\documentclass[aps,twocolumn,amsmath,samsmath,superscriptaddress,showpacs,floatfix]{revtex4}
\usepackage{epsfig}
\usepackage{graphicx}
\usepackage{amssymb}

\begin{document}
\title{Gel to glass transition in simulation of a
valence-limited colloidal system}

 \author{E. Zaccarelli}
 \affiliation{
  Dipartimento di Fisica and CNR-INFM-SOFT, Universit\`a di Roma `La
 Sapienza', P.le A. Moro~2, I-00185, Roma, Italy}
 \affiliation{ISC-CNR, Via dei Taurini 19, I-00185, Roma, Italy}

 \author{I. Saika-Voivod}
\affiliation{Department of Chemistry, University of Saskatchewan,
 Saskatoon, Saskatchewan, S7N 5C9}
 \affiliation{Dipartimento di Fisica,
 Universit\`a di Roma `La Sapienza', P.le Aldo
 Moro~2, I-00185, Roma, Italy}

 \author{S.~V. Buldyrev}
 \affiliation{
 Yeshiva University,  Department of
 Physics, 500 W 185th Street New York, NY 10033, USA}

\author{A.~J. Moreno}
\affiliation{
Dipartimento di Fisica and CNR-INFM-SMC, Universit\`a di Roma `La
Sapienza', P.le A. Moro~2, I-00185, Roma, Italy}
\affiliation{
Donostia International Physics Center,
Paseo Manuel de Lardizabal 4,
E-20018 San Sebasti\'an, Spain}
 \author{P. Tartaglia}
 \affiliation{
 Dipartimento di Fisica and CNR-INFM-SMC, Universit\`a di Roma `La
 Sapienza', P.le A. Moro~2, I-00185, Roma, Italy}

 \author{F. Sciortino}
 \affiliation{
 Dipartimento di Fisica and CNR-INFM-SOFT, Universit\`a di Roma `La
 Sapienza', P.le A. Moro~2, I-00185, Roma, Italy}

\date{\today}

\begin{abstract}
We numerically study a simple model for thermo-reversible colloidal
gelation in which particles can form reversible bonds with a
predefined maximum number of neighbors. We focus on three and four
maximally coordinated particles, since in these two cases the low
valency makes it possible to probe, in equilibrium, slow dynamics down
to very low temperatures $T$.  By studying a large region of $T$ and
packing fraction $\phi$ we are able to estimate both the location of
the liquid-gas phase separation spinodal and the locus of dynamic
arrest, where the system is trapped in a disordered non-ergodic
state. We find that there are two distinct arrest lines for the
system: a {\it glass} line at high packing fraction, and a {\it gel}
line at low $\phi$ and $T$. The former is rather vertical
($\phi$-controlled), while the latter is rather horizontal
($T$-controlled) in the $(\phi-T)$ plane.  Dynamics on approaching the
glass line along isotherms exhibit a power-law dependence on $\phi$,
while dynamics along isochores follow an activated (Arrhenius)
dependence.  
%We complement the molecular dynamics results with mode
%coupling theory calculations, using the numerical structure factors as
%input, and find that the theory satisfactorily predicts the glass line
%but fails in predicting the gel line. Indeed, 
The gel has clearly distinct properties from those of both a repulsive
and an attractive glass.  A gel to glass crossover occurs in a fairly
narrow range in $\phi$ along low $T$ isotherms, seen most strikingly
in the behavior of the non-ergodicity factor.  Interestingly, we
detect the presence of anomalous dynamics, such as subdiffusive
behavior for the mean squared displacement and logarithmic decay for
the density correlation functions in the region where the gel dynamics
interferes with the glass dynamics.

%% On lowering $T$, we obtain nearly perfect networks of bonded
%% particles, and observe nearly Arrhenius behavior along isochores of
%% the viscosity, diffusivity, bond lifetime and of the energy difference
%% from the ground state of the system.  Thus, at low $\phi$ we are able
%% to approach the gel state from equilibrium. W As $\phi$ increases,
%% $f_{\rm q}$ shows a rapid crossover from a gel to a hard-sphere like
%% glass.

\end{abstract}

\pacs{82.70.Gg, 82.70.Dd, 64.70.Pf }
% 82.70.Gg Gels and sols
% 61.20.Lc Structure of liquids; time-dependent properties; relaxation
% 61.43.Hv fractals, macroscopic aggregates including DLA
% 82.70.Dd Colloids
% 64.60.Ak Renorm. group, fractal, and percolation studies of phase tranistions
% 64.70.Pf Glass transitions

\maketitle
\section{Introduction}

Systems composed of mesoscopic solid particles dispersed in a fluid
are named colloids.  The properties of the particles and of the fluid
can be controlled via chemical or physical manipulations to a great
extent.  As a result the particle-particle interaction can be tuned
from very short range depletion attractions to very long range
Coulombic repulsion~\cite{Yeth03}, making colloids important both in
terms of basic scientific research and industrial
applications~\cite{Russellbook,Hansenbook,Fre02a,advances}.  The
canonical model system for colloids is the hard sphere system, for
which sterically stabilized colloidal particles such as PMMA provide a
very accurate experimental realization~\cite{vanM93,Bri02a}.  Hard
sphere colloids have been used to directly observe crystal nucleation
and to test theoretical predictions surrounding glass
transitions\cite{Pus87a}. The addition of small, non-adsorbing
polymers to a hard-sphere solution leads to a short-range effective
attraction between colloids through the so-called depletion
interaction~\cite{Asa58a,Lik01b}.  The size of the small polymer
controls the range of attraction, while the concentration controls the
strength of attraction $u_0$.  Neglecting the role of solvent
interactions, these systems can be simulated on a computer with a
short-range attractive potential, as simple as a hard core
complemented by a square well (SW).  Colloid-polymer mixtures have
been found to offer new scenarios of arrested states.  These hard
sphere plus short range attraction systems exhibit the usual hard
sphere glassy dynamics near $\phi \approx 0.6$.  However, a
fascinating phenomenon arises when the range of attraction is less
than approximately $10\%$ of the hard sphere diameter, that is a
re-entrance of the glass transition line, predicted by Mode Coupling
Theory (MCT) \cite{Fab99a,Ber99a,Daw00a}, and confirmed by several
experiments\cite{Mal00a,Pha02a,Eck02a,Chen03a,Pon03a,Gra04a} and
simulations \cite{Pue02a,Fof02a,Zac02a,Zac04a,Pue05a}.  In this case,
for a particular range of $\phi$, arrest can be achieved by either
increasing or decreasing $T/u_0$, the ratio of temperature $T$ to
attraction strength $u_0$.  The two types of glass are now commonly
named hard-sphere or repulsive glass for the one at higher $T$, and
attractive glass for the one at lower $T$.

At low densities short-range attractive colloids exhibit particle
clustering and gelation\cite{Asn97,Seg01a,Sha03c,Cip05a}. Recently,
several numerical works have focused on colloidal gelation
~\cite{Kum01a,Sci04a,Coni04,Sciobartlett,Zaccagel,DelG05}, with
the aim of better characterizing colloidal gels and attempting to
formally connect gel to glass formation
\cite{Seg01a,Ber99a,Ber03a,Kro04a,Sci04a}.  Following MCT ideas, the
gel state was interpreted as a low $\phi$ extension of the high $\phi$
attractive glass~\cite{Ber99a,Ber03a,Kro04a,Pue03a} and, generating
some confusion, the term "gel" is still often interchanged with the term
`attractive glass'.

Differently from chemical gelation, that was extensively studied in
polymer physics\cite{flory,Rub99a,Bro01a} and modeled in computer
simulations\cite{Liu97b,Ver01a,Pli03,Wen03,DelG03a,Voi04a}, colloidal
gelation is still quite poorly understood. Gelation arises when a
stable particle network is formed due to bonding. For chemical gels,
bonds are irreversible, and thus gelation can be explained in terms of
percolation theory. However, the bonds intervening in colloidal
aggregation have an energy typically of the order of $k_B T$, as for
example bonds induced by depletion interactions.  Thus such bonds are
mostly transient at finite attraction strength.  The existence of a
finite bond lifetime creates a gap between the location of the
percolation line and the dynamic arrest (gel) line.  Earlier
simulations on a lattice\cite{Liu97b,Gim01a,DelG03a} and
off-lattice\cite{Voi04a} have discussed the influence of reversible
bond formation on the gelation process.

The quest for bond stabilization often calls for exploring regions of
very high attraction strength in the phase diagram to increase the
bond lifetime and promote gelation. However, in this region, at low
packing fraction $\phi$ and at low $T$ (or at high $u_0$ values), a
phase separation into gas (colloid-poor) and liquid (colloid-rich)
always takes place\cite{And02a}.  Recent theoretical studies have
addressed the question of the relative location of the phase
separation and of the attractive glass line. It has been
found\cite{Zaccapri,Fof05a} that the attractive glass line meets the
spinodal at a finite temperature, on the high density side.  At low
$\phi$, arrest in (spherically interacting) short-range attractive
colloids occurs only as arrested phase separation, where the liquid
phase is glassy, and hence the phase separation cannot proceed
fully~\cite{Zaccapri}.
This scenario was found in numerical
simulations of the square well potential for a well width ranging from
$15\%$ to arbitrarily small values\cite{Emanuelab,Zaccapri,Fof05a}, as
well as in Lennard-Jones potential\cite{Sas00a}.
Gels as a result of interrupted phase separation have been identified
in experiments\cite{Man05a} and
simulations\cite{Soga99,Lod98a,Zaccapri}.  For extremely deep
quenches (at very low $T$) irreversible gelation may occur through
diffusion limited cluster
aggregation~\cite{vicsekbook,Car92a,Sci95a,Pou99a,Gim99a}.

Other mechanisms must be invoked for stabilization of
thermo-reversible gels and suppression of the spinodal
line\cite{genova}. A possibility is to consider the effect of residual
charges on the colloidal particles, that may produce a longe-range
repulsive barrier in the interparticle potential, thus efficiently
stabilizing the bonds and preventing the condensation of the liquid
phase. The emergence of an equilibrium cluster phase has been recently 
evidenced both in experiments\cite{Strad04,Bartlett04} and
simulations\cite{Sci04a,Coni04,Sciobartlett} on charged colloidal and
protein suspensions. A dynamical arrest transition then follows, driven
by electrostatic repulsions between the clusters \cite{Sci04a} or by
cluster branching and percolation \cite{Coni04,Sciobartlett}.

Very recently, forming a gel {\it in equilibrium}, without the
interference of phase separation, has been achieved by limiting the
number of attractive interactions (bonds) between nearest neighbors.
Following ideas first introduced by Speedy and
Debenedetti~\cite{Spe94,Spe96}, a model for thermo-reversible gelation
has been studied numerically\cite{Zaccagel}.  This model, in which
particles interact via a SW potential with the addition of a
geometrical constraint in the maximum number of bonds $n_{\rm max}$ a
particle can form, describes particles interacting via a hard-core
plus $n_{\rm max}$ randomly-located sticky spots\cite{Kal03a}. 
%Is the Nmax constrain really a geometrical one?  Also, 
%does the Nmax model really describe a system with random sticky spots, 
%or just approximate it?  If the sticky spots are fixed, though random, 
%is the configurational entropy the same as in the Nmax case?
In this
respect, the model retains the spherical symmetry but incorporates
features of associating liquids\cite{Wer84a}. Other possibilities to
limit the number of bonded pairs implementing geometrical constrains
and retaining a spherical pair potential can be found in
Ref.\cite{Hue02a,Hue02b}, while the same basic ideas have inspired
recent work on self-assembly of super-molecular/nano structures
\cite{Glo03a,vanW05}.  In the literature it is possible to find
several related models, also based on limited short-range attractions
in which the bonding constraint is imposed via angular degrees of
freedom, though the focus has not always been on their dynamic
properties\cite{Kol87a,Sea99c,Kern03, DelG05,pwm}.  In general,
ref~\cite{Zaccagel} puts forward the hypothesis that, only when a
restricted part of the colloidal surface is active in the formation
of attractive bonds, dynamic arrest at low $\phi$ can be observed in
the absence of a phase separation.  An experimental test of the
previous hypothesis will hopefully be provided by the new generation
of `patchy' colloids, or colloids with `sticky spots' \cite{Man03a},
which is about to be synthesized.

Understanding gelation at low densities in short-range attractive
models may also be relevant to the study of proteins, since they are
expected to belong to the class of short-range attractive interacting
systems\cite{Lom99a,Sea99c}.  Indeed, the formation of arrested
disordered states at low densities often interferes with
crystallization, and this is possibly one of the reasons why proteins are
often difficult to crystallize~\cite{Pia00a,Cac03a}.

In a recent Letter~\cite{Zaccagel}, we showed that the $n_{\rm max}$ model
(detailed in section II below) allows us to study thermo-reversible
gels.  We showed that the signatures of the gel state, as for example
the non-ergodicity factor $f_{\rm q}$, are quite distinct from those
of both the hard sphere and the attractive glass.  In the present
study, we explore the full-$\phi$ dependence of dynamics at low $T$
for both $n_{\rm max}=3$ and $n_{\rm max}=4$.
%and compare our results
%to the predictions of Mode Coupling Theory (MCT) for the same model.
We observe a transition over a small range in $\phi$ from a gel to a
repulsive glass.  In Section II, we give details of the model and
simulations.  Section III contains the results for the calculated
phase diagram and compares the relative location of the thermodynamic
and of the kinetic arrest lines.
%, including prediction for the ideal
%glass lines calculated according to MCT.  
We also report static and
dynamic correlation functions.  In Section IV we discuss results
and in Section V conclusions are drawn.

\section{Simulation Details}

We perform event-driven molecular dynamics simulations of $N=10000$
particles of mass $m=1$ with diameter $\sigma=1$ (setting the unit of
length) interacting via a limited-valency square well potential.
The depth of the well $u_0$ is fixed to $1$, and the width $\Delta$ of
the square well attraction is such that $\Delta/(\sigma+\Delta)=0.03$.
$T$ is measured in units of $u_0$, and the unit of time $t$ is $\sigma
(m/u_0)^{1/2}$. This system is a one-component version of the binary
mixture that has been extensively studied
previously~\cite{Zac02a,Zaccapri,Zac03a,Sci03a}.  In the following we
will use the acronym SW to indicate the $\Delta/(\sigma+\Delta)=0.03$
standard square well potential.  The limited-valency condition is
imposed by adding a bonding constraint.  The square well form can be
used to unambiguously define bonded particles, i.e., particles with
centers lying within $\sigma$ and $\sigma + \Delta$ of each other are
bonded.  
%% The valency constraint allows particles to only have up to
%% $n_{\rm max}$ bonds.  Once a particle has $n_{\rm max}$ bonded
%% particles, its interaction with all other particles is of a
%% hard-sphere type (with diameter $\sigma$).  
The interaction between two particles $i, j$, each having
less than $n_{\rm max}$ bonds to other particles, or between two particles
already bonded to each other, is thus given by a square-well potential,
\begin{equation}
%\vspace{-1 mm}
V_{ij}(r)=
\begin{cases}
        ~~\infty ~~~~~~~\hspace{0.8 mm} r<\sigma    \\
        -u_0  ~~~~~~~ \sigma<r<\Delta  \\
        ~~~0      ~~~~~~~~ r>\sigma+\Delta.
\end{cases}
\label{eq:model1}
\end{equation}
When $i$ and/or $j$ are already bonded to $n_{\rm max}$ neighbors,
then $V_{ij}(r)$ is simply a hard sphere (HS) interaction,
\begin{equation}
V_{ij}(r)=
\begin{cases}
        ~~\infty ~~~~~~~~r<\sigma    \\
        ~~~0      ~~~~~~~~\hspace{0.8 mm} r>\sigma.
\end{cases}
\label{eq:model2}
\end{equation}
The resulting Hamiltonian of the system has a many-body term
containing information of the existing bonds. Due to the fact that the
list of existing bonds is necessary at any instant of the simulation,
configurations are saved storing also the bond list. Moreover, all
simulations are started from high temperature configurations where all
particle overlaps are excluded within the attractive well distance
$\sigma+\Delta$.  In cases of multiplicity of possible bondings, such
as for example when a bond is broken for a particle that was
fulfilling the $n_{\rm max}$ allowed bonds and more than one neighbour
particle lie within its attractive well, a random neighbour, with less
than $n_{\rm max}$ bonds, is chosen to form the new bond.

The idea of constraining the number of square well bonds a particle
can form was introduced by Speedy and
Debenedetti~\cite{Spe94,Spe96}. In contrast to their original
version of the model, where triangular closed loops were not allowed,
in the present model no constraints on minimal bonded loops are
introduced. Our model can be considered as a realization of particles
with $n_{\rm max}$ randomly located `sticky spots'\cite{Kal03a}. 
%Again, is our model really a realization of random sticky spot 
%colloids?  Would random sticky colloids have the same network 
%degeneracy?  The same pressure?
The limited-valency properly defines the ground state of the systems,
corresponding to a potential energy per particle $-u_0 n_{\rm max}/2$.
This is achieved when every particle has $n_{\rm max}$ filled bonds.
We note that fully bonded clusters of finite size may occur.  The
smallest fully connected cluster sizes are: $4$ (tetrahedra) for
$n_{\rm max}=3$; $6$ (octahedra) for $n_{\rm max}=4$; and $12$
(icosohedra) for $n_{\rm max}=5$.  This introduces the intriguing
possibility of forming a hard sphere gas of such clusters at low
$\phi$ and $T$, as seen from the simulations. We study in depth the
cases $n_{\rm max}=3$ and $4$, for which we already know that there
exist significant differences in the location of the spinodal lines as
compared to the $n_{\rm max}=12$ (or standard SW) case.

For all state points simulated, we first equilibrate the system at
constant $T$ until the potential energy and pressure of the system
reach a steady state, and the MSD reaches diffusive behavior, i.e.,
during the equilibration time particles move on average at least one
particle diameter.  A subsequent constant energy simulation is used to
gather statistics for all reported quantities.

%% Interestingly enough, the use of the limited valency SW model
%% increases the region of stability with respect to crystallization for
%% the simulated state points. Indeed, nowithstanding the use of a
%% one-component system, we are able to monitor slow dynamics up to
%% $\phi=0.56$ for moderately low $T$.

%% %%  The
%% important constraint we impose on the particles is described as
%% follows.  Consider a particle $A$ with $n_{\rm max}$ particles bonded
%% to it.  The potential between $A$ and all other particles (i.e. those
%% not bonded to it) is merely a hard sphere interaction with diameter
%% $\sigma$.  The particle $A$ continues to interact with its bonded
%% particles through the square well.  Put in another way, if a particle
%% is fully bonded, incoming particles only see a hard sphere, and no
%% attractive well.  Of course, a bonded particle may leave its bond with
%% $A$, whereupon $A$ reverts to having a square well interaction with
%% all particles.

To estimate the equilibrium phase diagram, we calculate the gas-liquid
spinodal and the static percolation line.  The latter is defined as
the locus in $(\phi,T)$ such that $50\%$ of the configurations possess
a spanning, or percolating, cluster of bonded particles. 
%Two particles are considered bonded if their potential energy is $-u_0$.  
To characterize the structure and dynamics of the system, we evaluate:
the static structure factor,
\begin{equation}
S(q)\equiv \left< \left | \rho_q(0) \right |^2\right>,
\end{equation}
the mean squared displacement (MSD),
\begin{equation} 
<r^2(t)>\equiv <\sum_{i=1}^N \left |\vec r_i(t) - \vec r_i(0)\right
|^2/N>,
\end{equation}
the diffusion coefficient,
\begin{equation} 
D \equiv \lim_{t\rightarrow \infty} \left< \sum_{i=1}^N \left |\vec r_i(t) - \vec r_i(0) \right |^2\right>/6Nt,
\end{equation}
%% the viscosity $\eta = \left(2Vk_BT\right)^{-1}
%% \frac{\rm d}{{\rm d}s} \left< \left[A(s)-A(0)\right]^2 \right>$, where
%% $V$ is the volume of the system, $A(t)=m\sum_{i=1}^N \dot x_i y_i$,
%% $\dot x$ is the time derivative of $x$, and $x_i$ and $y_i$ are
%% orthogonal components of $\vec r_i$ (following the treatment given in
%% ~\cite{alder});
the dynamic structure factor, or density autocorrelation function,
\begin{equation} 
F_q(t)\equiv \left<\rho_q(t)\rho_{-q}(0)\right>
/\left< \rho_q(0)\rho_{-q}(0)\right>,
\end{equation}
and its long time limit or plateau value $f_q$, i.e. the
non-ergodicity parameter.  In all cases, $\left< . \right>$ denotes an
ensemble average, $\vec r_i$ is the position vector of a particle,
$\vec q$ is a wavevector and $i$ labels the $N$ particles of the
system, while $\rho_q(t)=\frac{1}{\sqrt{N}} \sum_{i=1}^{N}
\exp{(-i\vec q\cdot \vec r_i)}$.

Also, we monitor the bond lifetime correlation function, averaged over
different starting times, and defined as,
\begin{equation}
\phi_B(t)=\langle \sum_{i<j} n_{ij}(t)n_{ij}(0)\rangle / [N_B(0)],
\end{equation}
where $n_{ij}(t)$ is 1 if two particles are bonded up to time $t$ and
0 otherwise, while $N_B(0) \equiv \langle \sum_{i<j}n_{ij}(0) \rangle
$ is the number of bonds at $t=0$. We note that $\phi_B$ counts which fraction
of bonds found at time $t=0$ persists at time $t$, without ever breaking
within the store rate of configurations.  Associated with
$\phi_B(t)$, we extract an estimate of the bond lifetime $\tau_B$ via
stretched exponential fits.

%Questo 
%dettaglio non lo sapevo. Quale e l'store rate?
%Il tauB calcolato cosi sarebbe piu lungo che il tauB reale (quanto piu?). 
%Sappiamo che tauB(calc) > tauq, ma secondo tutto questo,
%e anche certo che tauB(reale) > tauq?.

Although colloidal systems are more properly modeled using Brownian
dynamics, we use event driven molecular dynamics due to its efficiency
in the case of step-wise potentials.  While the short-time dynamics is
strongly affected by the choice of the microscopic dynamics, the long
term structural phenomena, in particular close to dynamical arrest,
are rather insensitive to the microscopic dynamics~\cite{Gle98,pwm}.
To have a confirmation of this, we also performed additional
simulations where the effect of the solvent was mimicked by so-called
ghost particles \cite{Zhou97}.  In particular, we studied a system of
1000 colloidal particles and 10000 ghost particles and we found that
the long-time behavior of the dynamical quantities, such as the $T$
dependence of $\tau_B$ and the $q$ dependence of $f_q$ are independent
from the microscopic dynamics.  We also note that equivalence between
Newtonian and Brownian dynamics is not guaranteed when studying small
length scales (as for example the decay of density fluctuations at
large $q$), i.e. for intra cage motion, since there the microscopic
dynamics may affect the shape of the decay of the correlation
function. This is most relevant in gel systems for which the cage
length can be significantly larger than the particle size.

\section{Results}

In this section we examine the results of our simulation in terms of
the thermodynamic and dynamic quantities mentioned above.  We focus
our attention on the cases $n_{\rm max}=3$ and $n_{\rm max}=4$, where
a significant suppression of the liquid-gas spinodal as compared to
the SW case is observed\cite{Zaccagel}.  The suppression of the
critical temperature and the shrinking of the unstable region in the
$(\phi-T)$ plane makes it possible to study in one-phase conditions
state points characterized by an extremely slow dynamics.  More
precisely, it is possible to study without encountering phase
separation all $\phi \gtrsim 0.2$ for $n_{\rm max}=3$ and $\phi
\gtrsim 0.30$ for $n_{\rm max}=4$. In the SW case, phase separation
was encountered already at $T \approx 0.32$, when particle mobility is
always large\cite{Zaccapri}.

\subsection{Phase Diagram}

Ref.~\cite{Zaccagel,genova} show that constraining the number of
bonded neighbors reduces the energetic driving force for particle
clustering. 
%Is it Ref.[] show or shows?  How does the journal make a plural out of Ref.?
Therefore, as $n_{\rm max}$ decreases, the phase
separation transition (that we monitor by studying its spinodal), and
the percolation line both shift to lower $T$ (along isochores) and
lower $\phi$ (along isotherms).

\begin{figure}
\hbox to\hsize{\epsfxsize=0.9\hsize\hfil\epsfbox{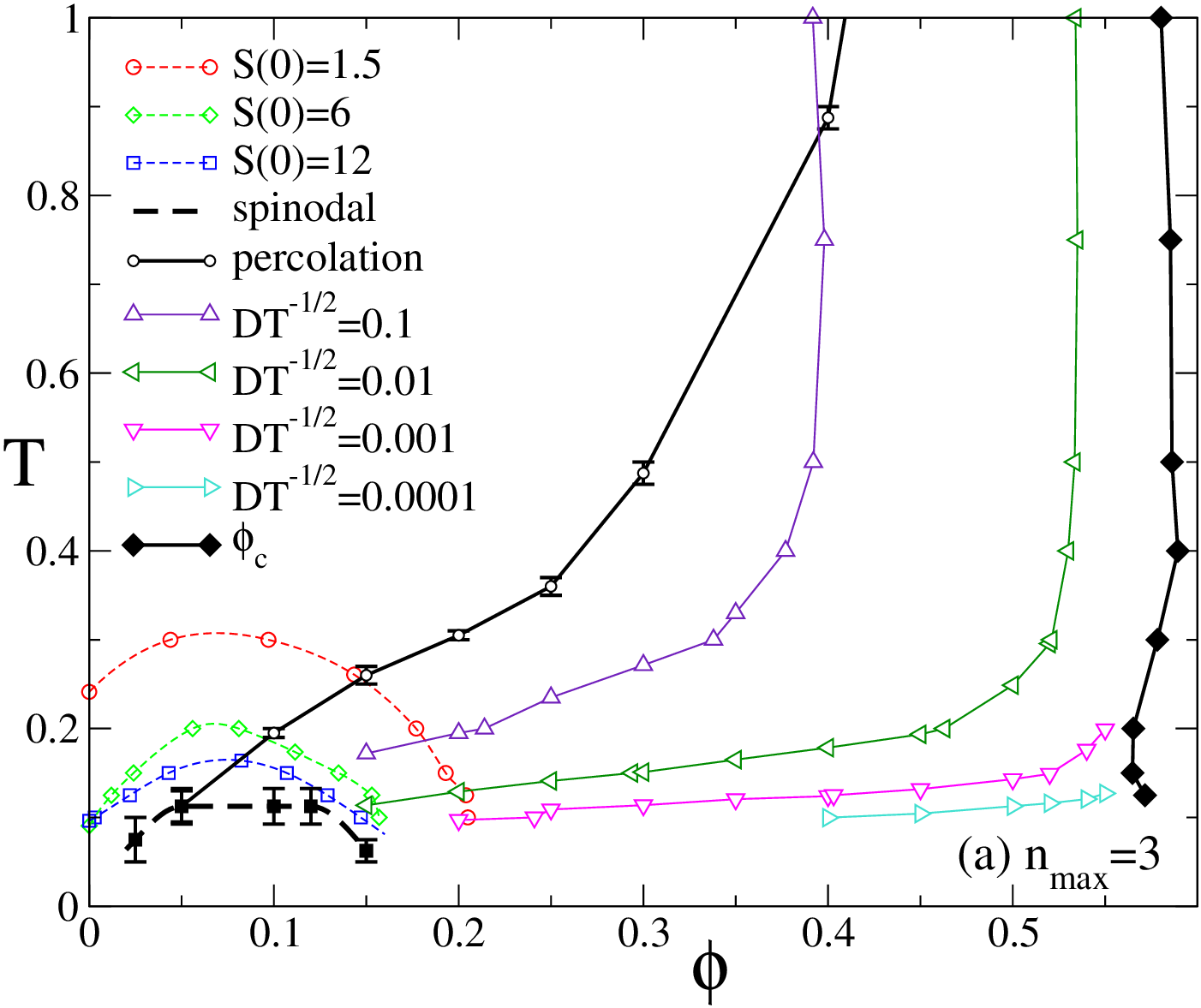}
\hfil}
\vspace{0.35cm}
\hbox to\hsize{\epsfxsize=0.9\hsize\hfil\epsfbox{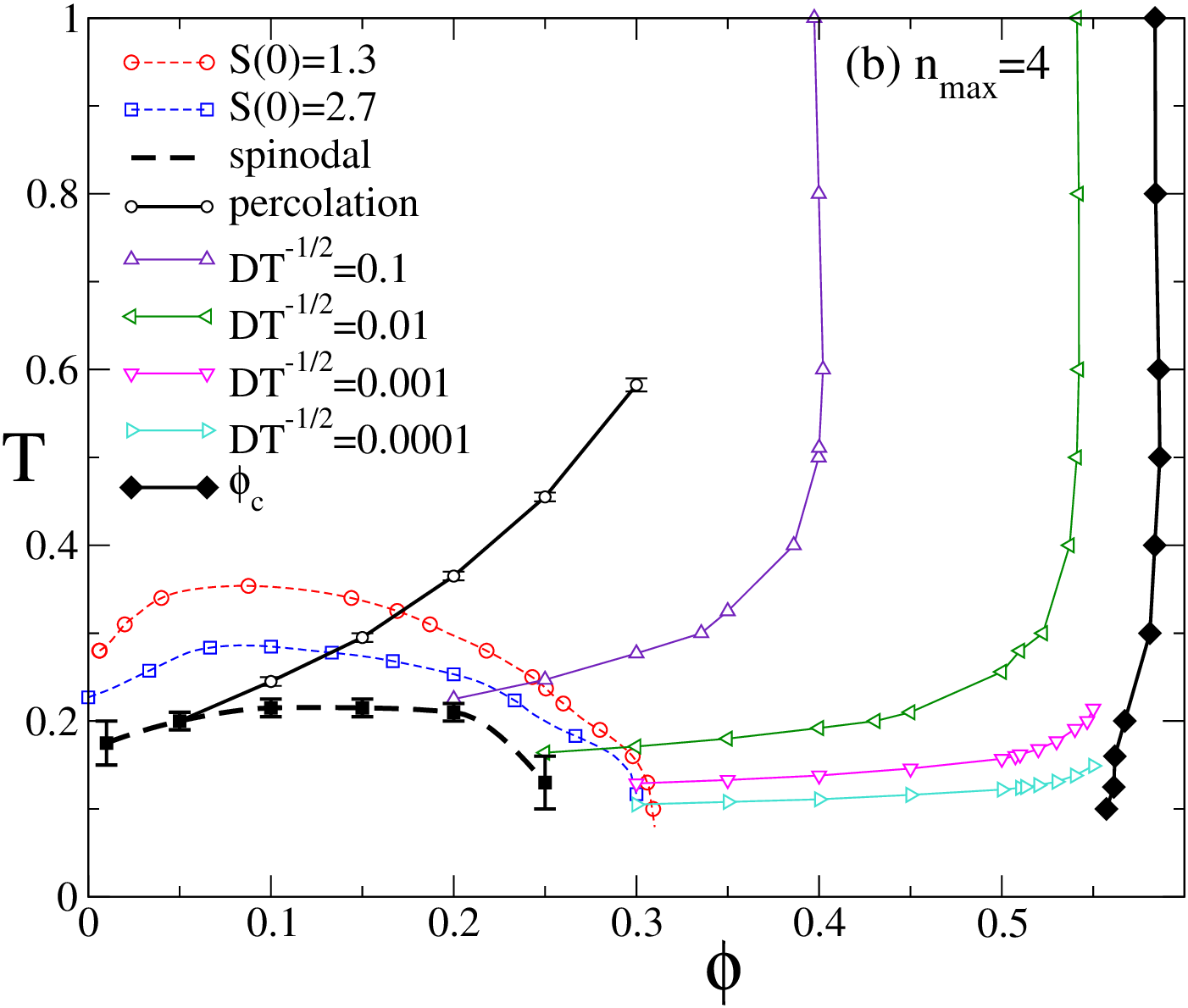}\hfil}
\caption{ Phase diagram for $n_{\rm max}=3$ (a) and $n_{\rm max}=4$
(b), showing the spinodal (dashed lines with squares), percolation
(solid lines with open circles), iso-diffusivity loci where
$D/\sqrt{T}= constant$ (lines with triangles), `iso-$S(0)$ lines'
(dashed lines).  Also shown is the
extrapolated {\it glass} line, labeled as $\phi_c$
%, and {\it gel}, labeled as $T_c$, lines 
from power-law fits, see text.
}
\label{fig:perc_spin}
\end{figure}

Fig.~\ref{fig:perc_spin} reports the percolation and spinodal loci for
$n_{\rm max}=3$ (a) and $n_{\rm max}=4$ (b).  To evaluate the location
of the spinodal line we interpolate the pressure $P(\phi)$ and search
for the condition $dP/d\phi=0$.  To better track down the spinodal and
estimate the effect of the spinodal in its vicinity we also determine
the loci of constant $S(q\rightarrow 0)$, that we name `iso-$S(0)$
lines', that can be considered as precursors of the spinodal line.
Indeed, $S(0)$ is connected to the isothermal compressibility
$\kappa_T=(d\phi/dP)/\phi$ by $S(0)=\rho k_B T \kappa_T$.  The
iso-$S(0)$ lines are calculated as loci of constant $(dP/d\phi)/k_B
T$.  We cross-check these estimates with the less precise value
obtained calculating directly $S(q\rightarrow 0)$.  The iso-$S(0)$ lines
are shown to emphasize that, in bonded systems, an increase at
small-$q$ in the scattering intensity can arise in the one-phase
region due to the vicinity of the spinodal curve.

Fig.~\ref{fig:perc_spin} also shows iso-diffusivity lines, i.e. lines
where $D/\sqrt{T}$ is constant. The scaling factor $\sqrt{T}$ is used
to take into account the trivial contribution of the thermal velocity
with $T$.  The investigated values of $D/\sqrt{T}$ cover four
orders of magnitude.  Such iso-diffusivity lines are precursors of the
dynamical arrest transition, corresponding to $D \rightarrow
0$. Previous works \cite{Fof02a,Zac02a,Fof03a} have shown that these
lines provide estimates of the shape and location
%(via some extrapolation, e.g. power law fits)
of the arrest line.  The isodiffusivity lines show an interesting
behavior. They start from the high-$\phi$ side of the spinodal curve
and then end up tracking the high-$T$ hard-sphere limit. They are
rather horizontal (parallel to the $\phi$-axis) at low $\phi$ and
rapidly cross to a vertical shape (parallel to the $T$-axis) at high
$\phi$.  The crossing from horizontal to vertical becomes sharper and
sharper on decreasing $D/\sqrt{T}$.  In the SW case, the
iso-diffusivity lines exhibit a reentrance in $\phi$, in agreement
with the predictions of the Mode Coupling Theory (MCT). The reentrance
becomes more and more pronounced at lower and lower $D/\sqrt{T}$
values.  In the $n_{\rm max}$ case, a reentrant shape is hardly
observed.  Indeed, in the SW the reentrance arises from the
competition between cages created by the nearest neighbors excluded
volume, with a typical hard-sphere localization length $\sim 0.1
\sigma$, and cages created by bonding with a localization length $\sim
\Delta$. In the $n_{\rm max}$ case, such a competition becomes less
effective due to the smaller number of bonds.

\subsection{Estimation of dynamical arrest lines}

 \begin{figure}
 \hbox to \hsize{\epsfxsize=0.75\hsize\hfil\epsfbox{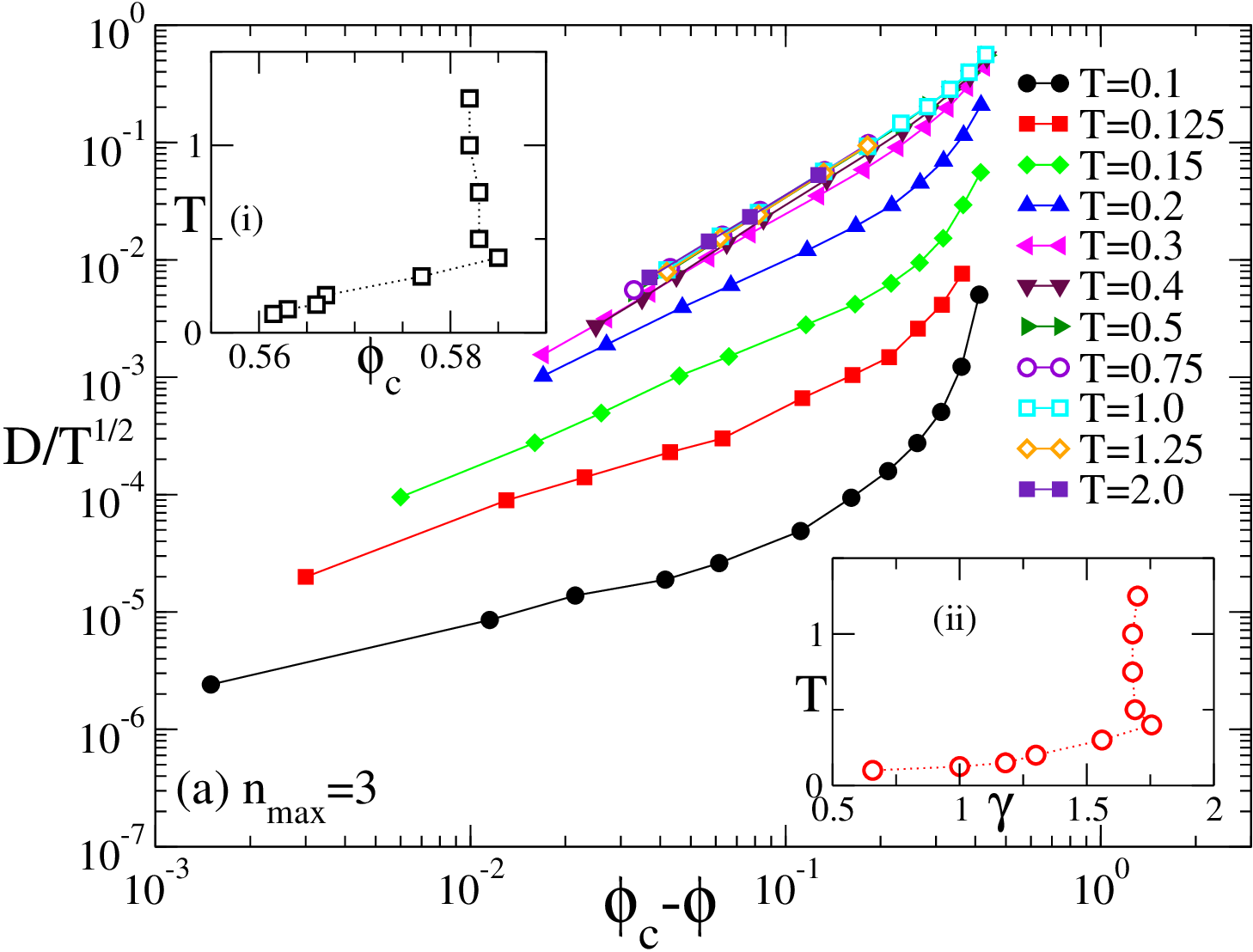}}
\vspace{0.5cm}
 \hbox to \hsize{\epsfxsize=0.75\hsize\hfil\epsfbox{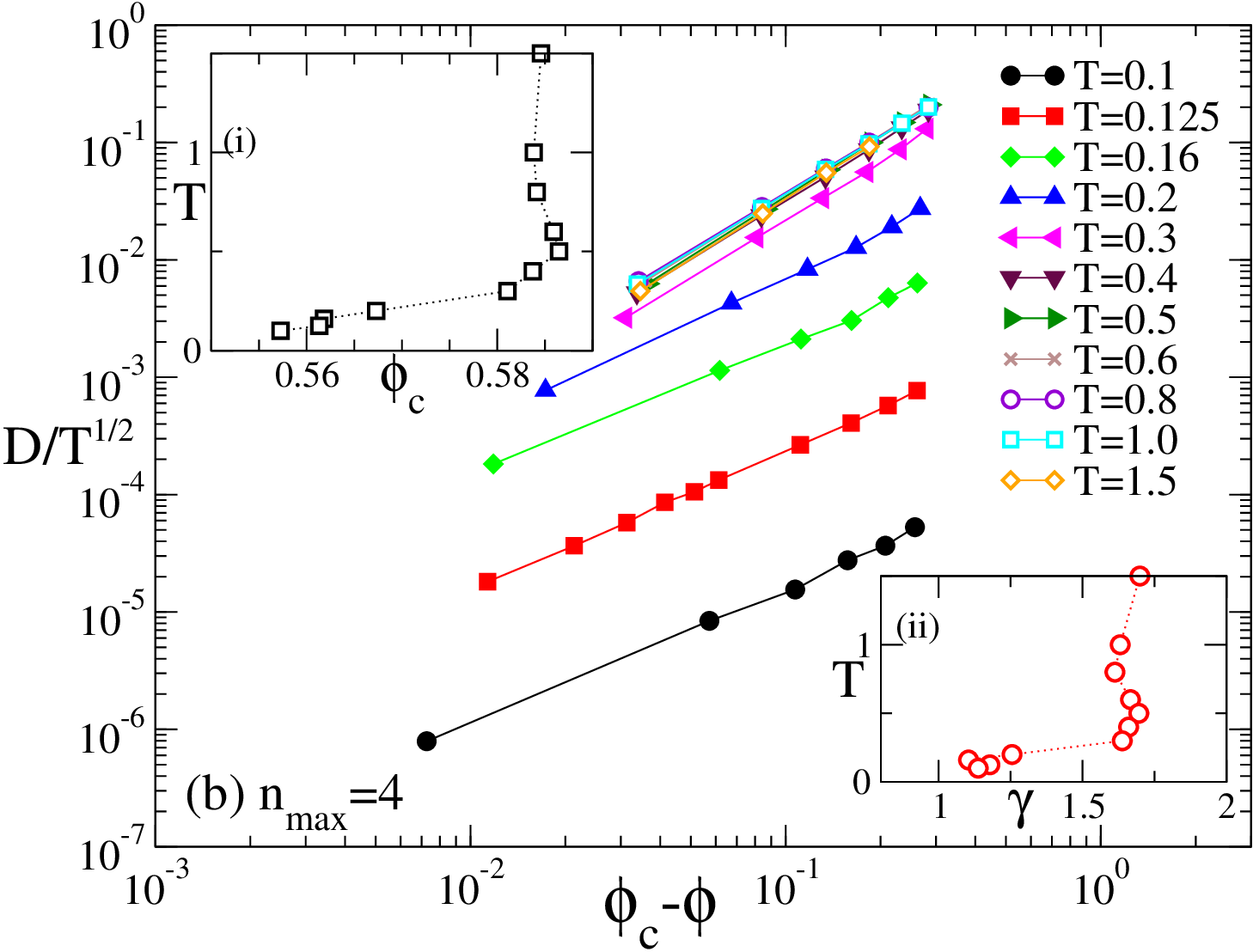}}
 \caption{ Power law fit of inverse diffusivity $D^{-1}=A(\phi_c -
 \phi)^{-\gamma(T)}$ for (a) $n_{\rm max}=3$ and (b) $n_{\rm max}=4$.
 Lines are guides to the eye.  Insets (a-i) and (a-ii) show the
 behavior of $\phi_c$ in the $\phi-T$ plane.  Inset (b-ii) and (b-ii)
 show the exponent $\gamma$ as a function of $T$.  }
 \label{fig:power}
 \end{figure}

To provide an estimate of the dynamical arrest lines, we can identify
a range of parameters where the characteristic time follow a power-law
dependence.  In the case $D$ is selected, data can be fitted according
to $D^{-1}(\phi)=A(\phi_c - \phi)^{-\gamma(T)}$, where $\phi_c$ is the
best estimate for the glass line.  Fig.\ref{fig:power} shows in
log-log scale $D^{-1}(\phi)/\sqrt{T}$ vs $(\phi_c-\phi)$ for $ n_{\rm
max}=3$ and $ n_{\rm max}=4$, where the $\phi_c$ values are chosen by
a best-fit procedure.  For $(\phi_c-\phi)\lesssim 0.3$, data are found
to be well-represented by power-laws for at most two decades in $D$.
The fits are performed over the range $(\phi_c-\phi)<0.3$.  The $T$
dependence of the fit parameters is also reported in
Fig.\ref{fig:power}. The two fit parameters vary almost in phase with
each other for both $n_{\rm max}$ values.  At high temperatures ($T
\gtrsim 0.3$) attraction does not play a role, the arrest line is
almost vertical and $\phi_c$ and $\gamma$ are practically constant.
For $T\lesssim0.3$ smaller values of $\phi_c$ and $\gamma$ are found.
The values of $\gamma$ are rather small at high $T$, close to the
lowest possible value allowed by MCT, and become smaller than the
lowest possible value allowed by MCT at low $T$.
We note on passing that crystallization limits the $\phi$ range over
which dynamic measurements in (metastable) `equilibrium' can be
performed.  The region where we detect crystallization varies with
$T$.  At high $T$ crystallization happens already for $\phi>0.54$,
while at low $T$, crystallization does not intervene up to
$\phi=0.56$. Thus, we cannot fully rely on the high $T$ fits as
$(\phi-\phi_c)$ is always large. Notwithstanding this, the estimate
$\phi_c\approx 0.58$ is reasonable.  On the other hand, at low $T$,
the vicinity to $\phi_c$ increases, 
but the range of D values over which a power-law can be 
fitted decreases to about one decade, making the fit questionable.
The fact that $\gamma$  decreases well below the lowest meaningful MCT 
value could tentatively be associated to a
difficulty of the theory to handle the crossing to an energetic caging
(see also below).
The resulting
$\phi_c(T)$ line is also reported in Fig.~\ref{fig:perc_spin}.
Independently of the fitting procedure, a clear vertically shaped
arrest line, driven mostly by packing, is observed on isothermal
compression.

\begin{figure}
\hbox to \hsize{\epsfxsize=0.75\hsize\hfil\epsfbox{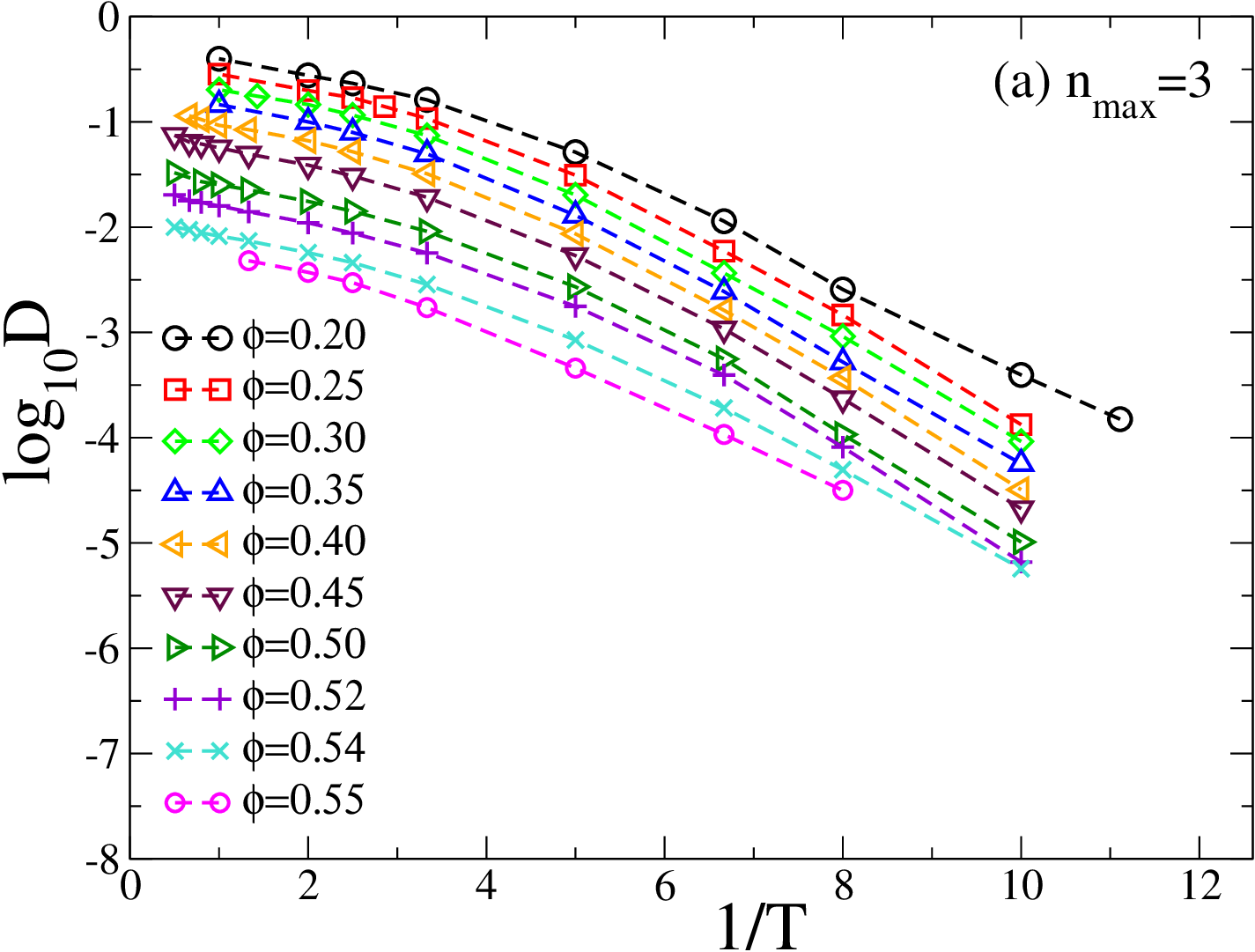}}
\vspace{0.6cm}
\hbox to \hsize{\epsfxsize=0.75\hsize\hfil\epsfbox{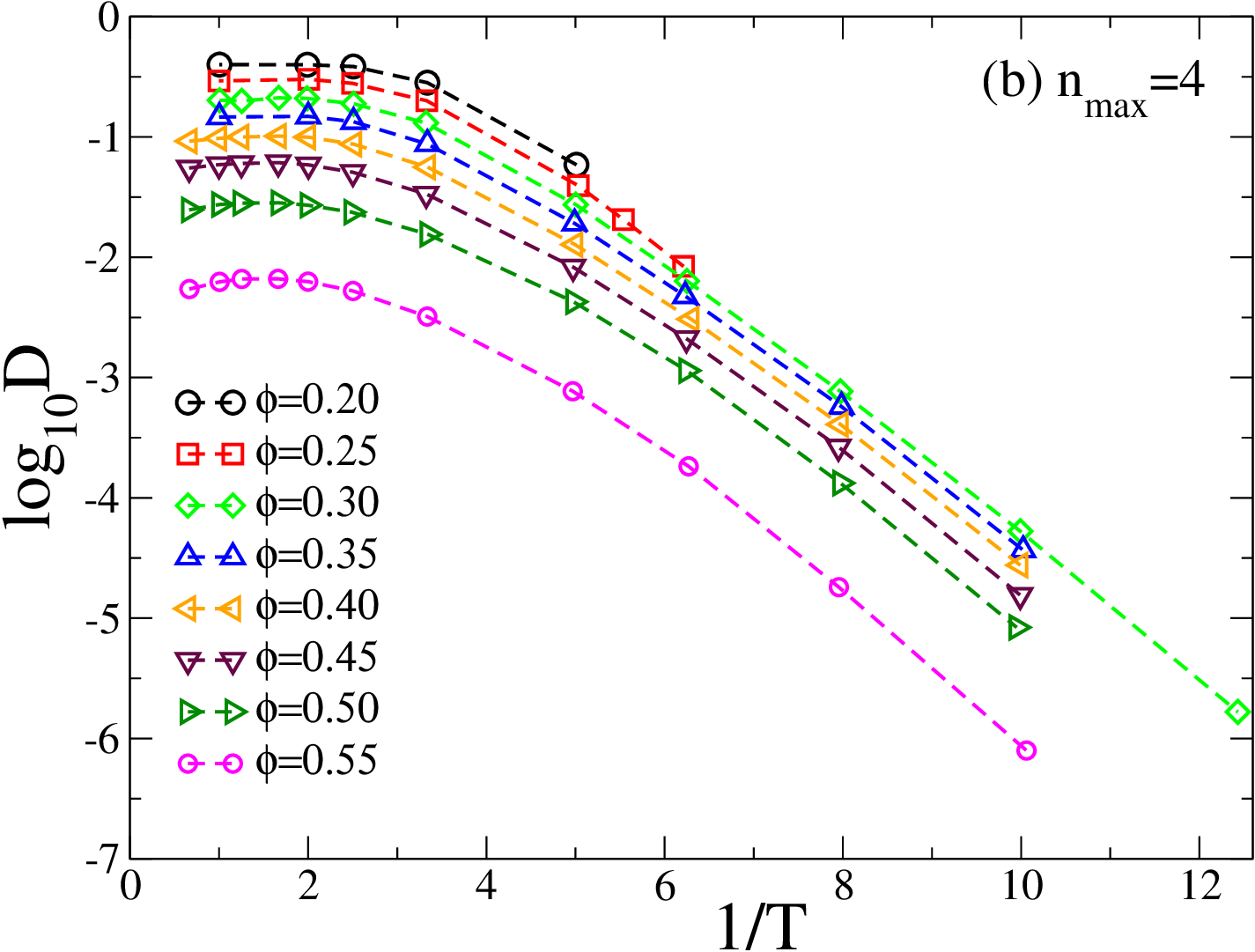}}
\caption{
Arrhenius plot of $D$ for $n_{\rm max}=3 $ (a) and $4$ (b) 
along all studied isochores
without intervening phase separation.
}
\label{fig:arrhenius}
\end{figure}

To provide a better estimate of the arrest line at low $\phi$ we have 
studied the behavior of $D$ with $T$ along isochores. We
have tried two routes.  The first one consists in performing again
power law fits but in temperature along an isochore, i.e.  $D(T)\sim
A(T -T_c)^{\gamma_T}$, selecting an appropriate $T$-interval. For both
$n_{\rm max}=3$ and $4$, such fits appear to hold for a rather small
interval in $D$ and are strongly dependent on the chosen $T$-interval
selected in the fit procedure. In this way, we cannot extract a
reliable estimate for $T_c$. Indeed, for the same state point $T_c$ could
vary from $0.2$ to $0.08$ depending on the fitting interval. However,
fixing a $T$-interval of fitting for all isochores, the resulting
$T_c(\phi)$ is again rather flat for both $n_{\rm max}$ values.
% The exponents $\gamma_T$ are too small
%(almost close to zero) in an intermediate $T$-regime, 
%suggesting 
%while the
%`ideal' critical temperature is $T \approx 0.08$ for all densities and
%both $n_{\rm max}$.  
The second route is more robust and is based on the observed low $T$
behavior, where data are found to follow very closely an Arrhenius
law. Fig.~\ref{fig:arrhenius} shows $D$ as a function of $1/T$, for
both $n_{\rm max}=3$ and $4$.  Arrhenius behavior of $D$ is observed
at all $\phi$ at low $T$.  The activation energy is
around $0.45$ for $n_{\rm max}=3$ and of $0.55$ for $n_{\rm max}=4$.
Since at the lowest studied temperatures the structure of the system
is already essentially $T$-independent (see later on the discussion
concerning Fig.\ref{fig:sq}), there is no reason to expect a
change in the functional law describing the $T \rightarrow 0 $
dynamics.  In this respect, the true arrest of the dynamics is located
along the $T=0$ line, limited at low $\phi$ by the spinodal and at
high $\phi$ by crossing of the repulsive glass transition line.  This
peculiar behavior is possible only in the presence of limited valency,
since when such a constraint is not present, phase separation preempts
the possibility of accessing the $T \rightarrow 0 $ Arrhenius window.

\begin{figure}
\hbox to\hsize{\epsfxsize=0.9\hsize\hfil\epsfbox{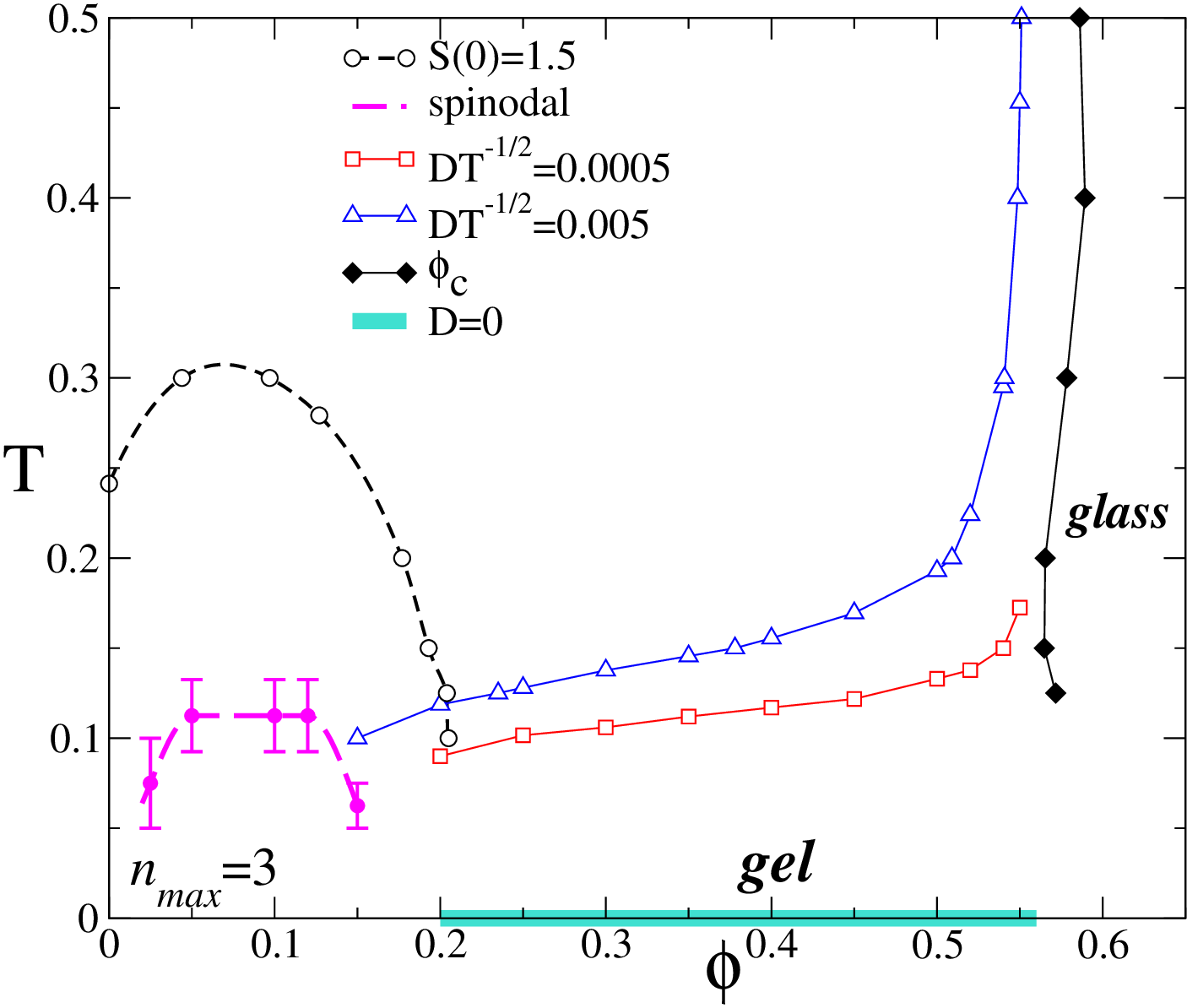}
\hfil}
\caption{Summary of the thermodynamic and kinetic phase diagram for 
$n_{\rm max}=3$, 
including spinodal (dashed lines with filled circles), iso-diffusivity
loci where $D/\sqrt{T}= 0.005, 0.0005$ (lines with triangles and
squares), iso-$S(0)$ locus where $S(0)=1.5$, 
%MCT lines (attractive and repulsive), 
extrapolated {\it glass}, labeled 
as $\phi_c$, and
{\it gel}, labeled as $D=0$, lines respectively 
from power law and Arrhenius fits
($T_c=0$), see text.   }
\label{fig:summary}
\end{figure}

We can summarize the dynamical arrest behavior in
Fig.~\ref{fig:summary}.  One locus of arrest is found at high $\phi$,
rather vertical and corresponding to the hard-sphere glass transition.
%This locus is quite well described by MCT.  Very different is the
%situation concerning the low $T$ slowing down.  
The isodiffusivity lines suggest a rather flat arrest line. Two
different loci could be associated to arrest at low $T$. One defined
by the $T_c$ of the power-law fits and one at $T=0$ associated to the
vanishing of $D$ according to the Arrhenius law.  Notwithstanding the
problem with the power-law fits and the big undeterminacy on $T_c$, it
would be tempting to associate the $T_c$-line to the attractive glass
line, at least as a continuation of it at low $\phi$, and
%predicted by MCT and 
interpret the wide region between the two lines
as a region of activated bond-breaking processes\cite{Zac03a}.  
Data reported in the next sections will show that, in the present model,
the identification of such line with an  
%MCT 
attractive glass line (or its extension to low density) is not valid,
independently of the fit results. Indeed, we will show that particles
are never localized within the attractive well width $\Delta$, at any
$\phi$. On the other hand,
% and that the signatures of the arrest are very
%different from the MCT predictions. We associate this non-MCT behavior
the establishment of a percolating network of long-lived bonds, that
we will refer to as a {\it gel} is identified.  Ref.  \cite{Zaccagel}
shows that arrested states at low $\phi$ and $T$ are profoundly
different from both attractive and hard-sphere glasses. Now
%that MCT predictions are available, it is possible 
we are going to investigate if on increasing $\phi$ there is a
crossover or a transition from gel to attractive glass, or whether the
attractive glass exists at all in limited-valency models.  We refer
the reader to a future work to compare these results with
corresponding MCT predictions for the same model \cite{CHEN}.

\subsection{Static structure factor}

This section reports results for the static structure factors for
various studied $T$ and $\phi$ and both $n_{\rm max}=3$ and $n_{\rm
max}=4$. Results along an isochore and an isotherm are general for
both studied $n_{\rm max}$ values.

\begin{figure}
\hbox to\hsize{\epsfxsize=0.9\hsize\hfil\epsfbox{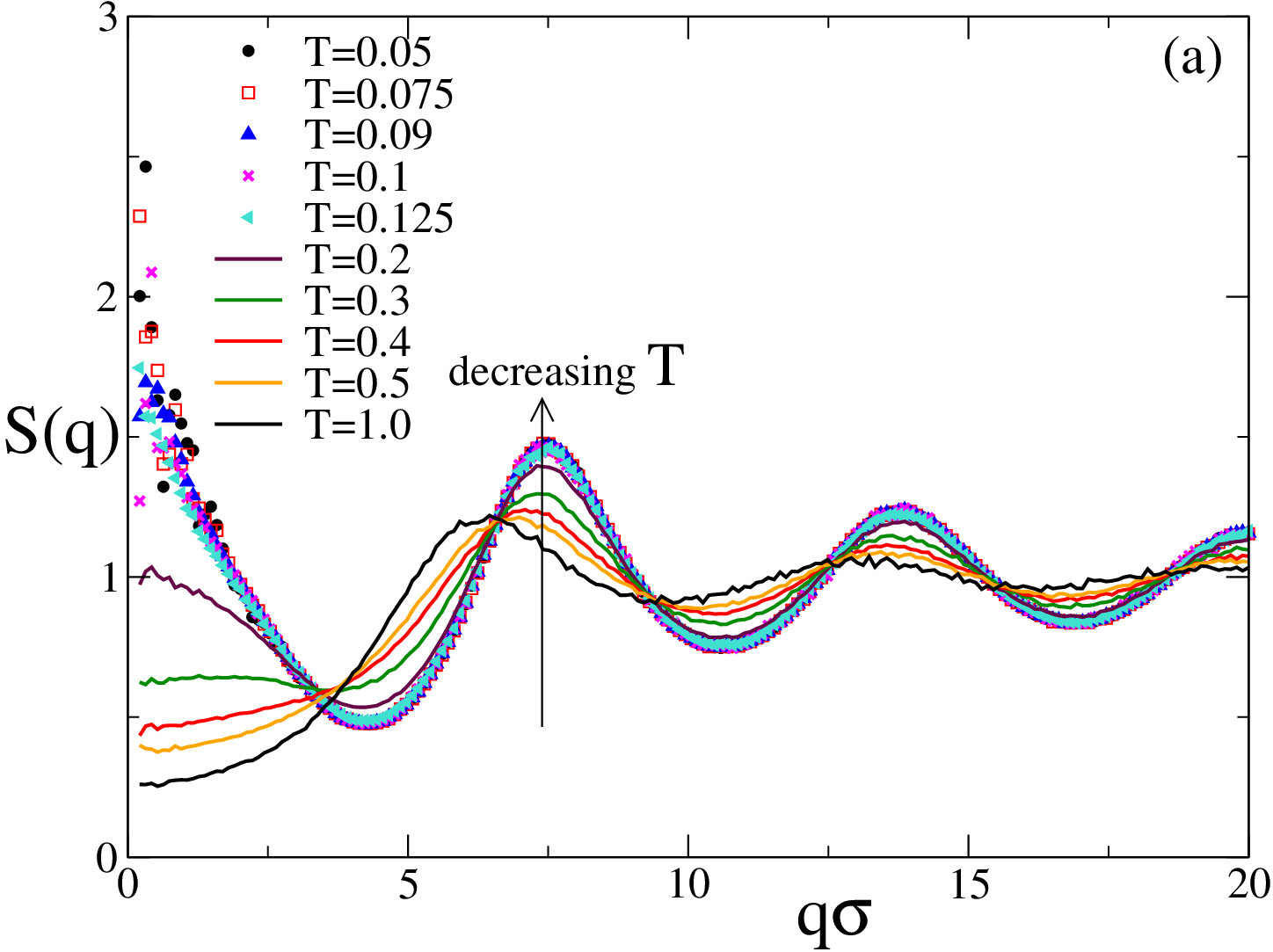}\hfil}
\vspace{0.6cm}
\hbox to\hsize{\epsfxsize=0.9\hsize\hfil\epsfbox{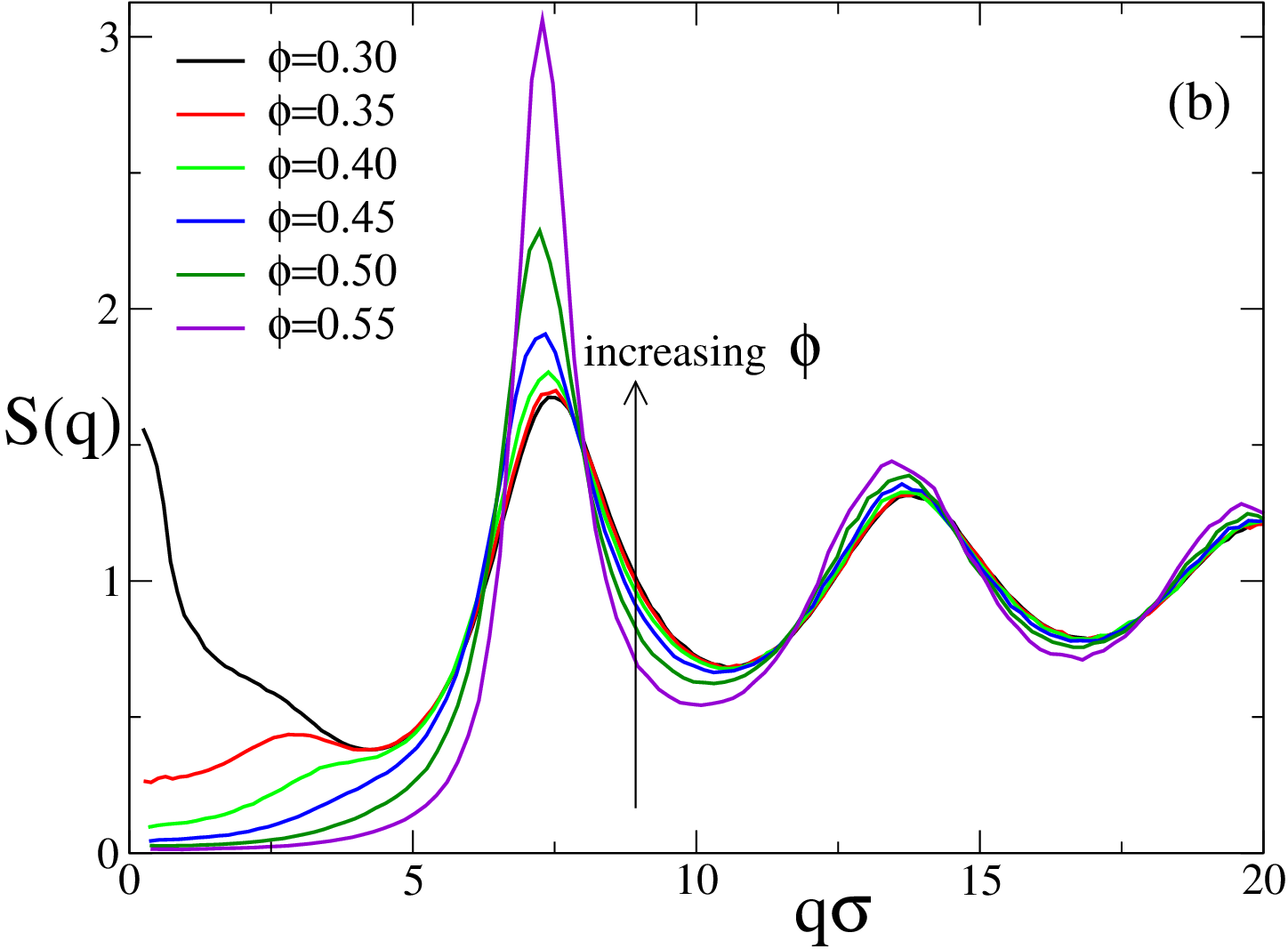}\hfil}
\caption{ (a) Evolution of the static structure factor $S(q)$ with $T$ for
the $\phi=0.20$ isochore for $n_{\rm max}=3$.  Below $T=0.125$ the
system has reached an almost fully connected state and $S(q)$ does not
change any longer with $T$;
(b) Evolution of the 
static structure factor $S(q)$ with  $\phi$ for the $T=0.125$
isotherm for $n_{\rm max}=4$. 
}
\label{fig:sq}
\end{figure}

Fig.~\ref{fig:sq}a shows the evolution with temperature of $S(q)$ at
the lowest accessible $\phi$ (i.e. the lowest $\phi$ where phase
separation is not present).  On lowering $T$, $S(q)$ shows an increase
of the intensity at small wave vectors, which saturates to a constant
value when most of the bonds have been formed. This indicates that the
system becomes more and more compressible, with large inhomogeneities,
characterizing the {\it equilibrium} structure of the system. The
inhomogeneities can be seen as an echo of the nearby phase separation
or, equivalently, as a consequence of building up a fully connected
network of particles with low coordination number.  The large signal
at small $q$ is a feature of $S(q)$ which is often observed in gel
samples \cite{Sha03c,Tan04aPRE}.  However, sometimes it may be difficult to
discriminate between a true equilibrium gel and an arrested state
generated through spinodal decomposition.  In the present model,
where phase separation is confined to the low $\phi$ region of the
phase diagram, it is possible to reach in equilibrium extremely low
$T$, making it possible to study reversible gel formation.

Besides the low-$q$ growth, on cooling $S(q)$ shows a progressive
structuring of peaks at $q \sigma \sim 2\pi$ and multiples thereof,
signaling the fact that particles progressively become more and more
correlated through bond formation. Indeed, the potential energy of the
system progressively approaches the ground state value, where all
particles have $n_{\rm max}$ bonds \cite{Moreno2005}.
Fig.~\ref{fig:sq}b shows the evolution of $S(q)$ on increasing $\phi$
along a low $T$ isotherm.  Moving further from the spinodal, the $q
\rightarrow 0$ peak decreases. A small pre-peak, around $q\sigma
\approx 3$ is present at low densities, and persists with smaller
intensity also at intermediate $\phi$. Beyond $\phi\simeq 0.45$, a
significant growth of the nearest-neighbour peak is found, signaling
the increasing role of packing.

\subsection{Mean squared displacement and caging}

One of the hallmarks of glassy dynamics is the caging of a particle by
its immediate neighbors.  Caging is most easily seen in a log-log plot
of the the mean squared displacement (MSD) vs time as an
intermediate-time plateau, separating short-time ballistic intracage
motion and long-time diffusion out of the cage. The height of the
plateau in the MSD provides a typical (squared) value for the
localization length $l_0$ of the particles within the arrested
state. For the standard HS glass, $l_0$ is found to be roughly
$0.1\sigma$, and corresponds to the average distance a particle can
explore rattling within its nearest neighbour cage. For an attractive
glass, on the other hand, $l_0$ corresponds to the attractive
well-width, since particles are forced to rattle within the bond
distance. For this reason, the MSD plateau is significantly smaller
than for the HS glass (of the order $\Delta^2 \approx 10^{-3}$ versus
$(0.1\sigma)^2\approx 10^{-2}$). For the same reason, the $q$-width of
$f_q$ is significantly larger for the attractive glass solution than
for the repulsive one.

In the low-$\phi$ study reported in \cite{Zaccagel}, on isochoric
cooling a clear plateau develops in the MSD, but its value indicates a
very high localization length, of the order of one or more particle
diameters.  The localization length does not change appreciably with
$T$, even below $T_c$.  We attributed this finding to the presence of
a long-living percolating network of particles allowing for ample
single particle movements (arising from the gel "vibrational" modes)
which completely mask the bond localization.  In this respect, the
connectivity of the network plays an important role in the slowing
down and provides an additional mechanism of arrest with respect to
both attractive and repulsive glasses.

\begin{figure}
\hbox to\hsize{\epsfxsize=0.9\hsize\hfil\epsfbox{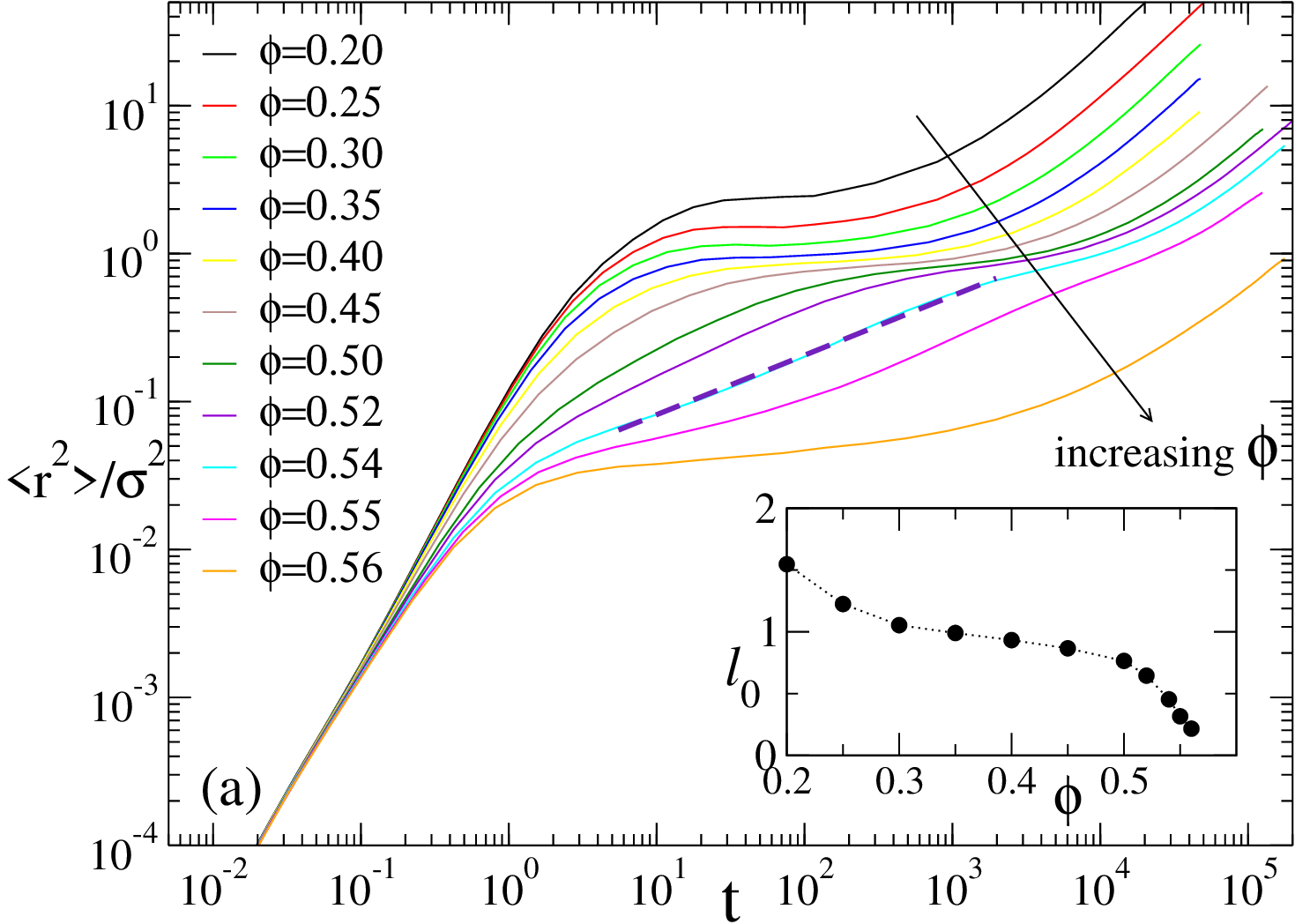}\hfil}
\vspace{0.5cm}
\hbox to\hsize{\epsfxsize=0.9\hsize\hfil\epsfbox{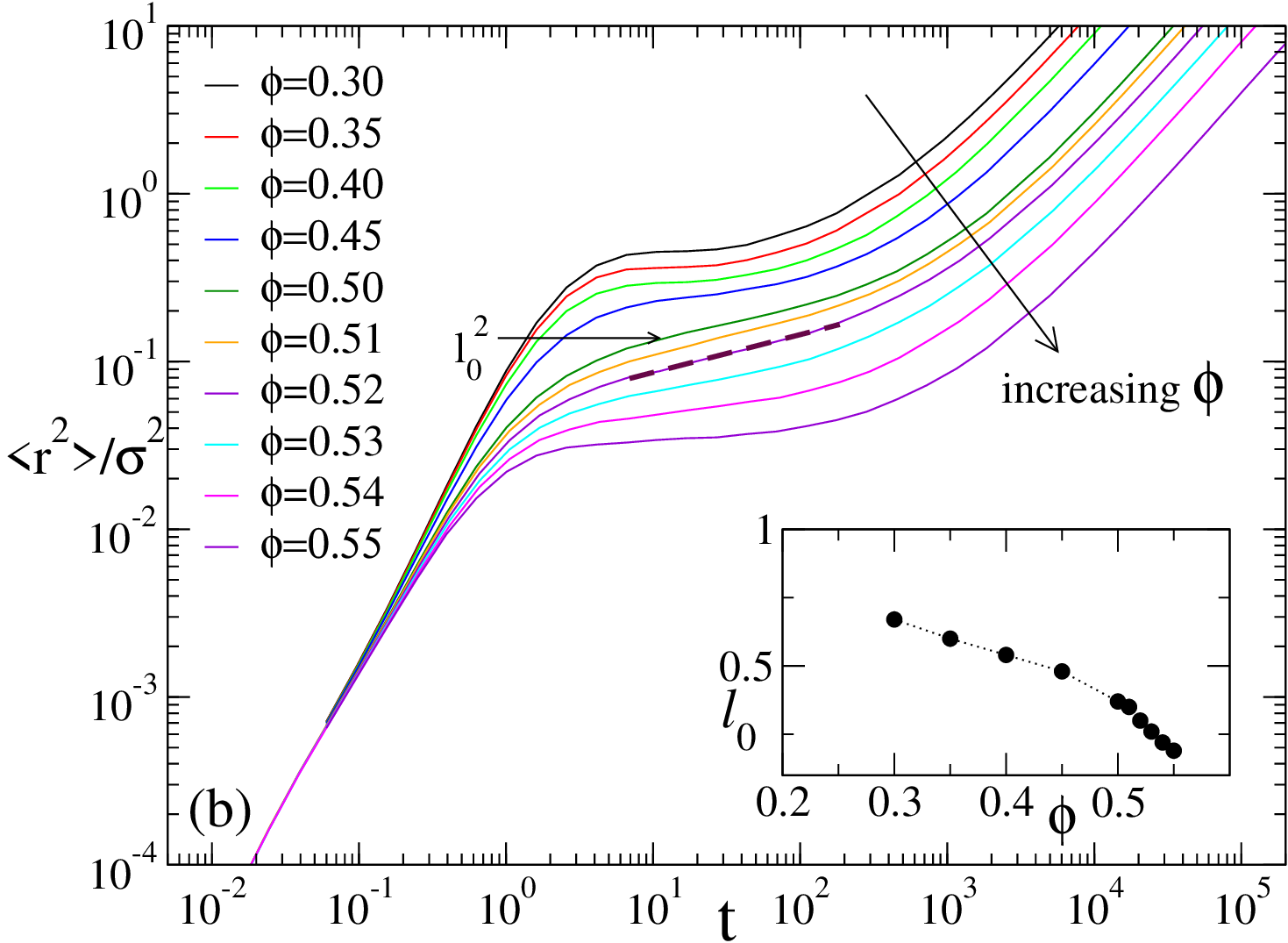}\hfil}
\caption{Mean squared displacement (MSD) and caging.  (a) MSD along
$T=0.1$ for $n_{\rm max}=3$; (b) MSD along $T=0.125$ for $n_{\rm
max}=4$.  The insets show the localization length $l_0$ as a function
of $\phi$, with dotted line as guide to the eye.
Dashed lines highlight subdiffusive
behavior at intermediate $\phi$. }
\label{fig:caging}
\end{figure}

Here we study the high-$\phi$ behavior with the aim of locating the
cross-over from low-$\phi$ arrest (gel) to the high-$\phi$ case and to
see if a crossover or transition emerges to one of the above cited
glasses.  Results for the $\phi$ dependence of the MSD along a low $T$
isotherm are presented in Fig.~\ref{fig:caging} for $n_{\rm max}=3$
and $4$.  Both graphs show similar features. A very high plateau of
order unity, slightly decreasing with $\phi$, is found up to
$\phi\approx 0.45$.  Although the long-time dynamics is monotonically
slower with increasing $\phi$, the plateau becomes less defined near
$\phi=0.50$, slowly crossing over to a quite distinct plateau
compatible with the HS one at $\phi=0.55$ for $n_{\rm max}=4$ and at
$\phi=0.56$ for $n_{\rm max}=3$.

Defining the cage length $l_0$ as the square root of the MSD value at
the inflection point of the MSD in log-log scale we plot $l_0$ in the
insets of Fig.~\ref{fig:caging}.  The cage length starts from values
larger ($n_{\rm max}=3$) or close ($n_{\rm max}=4$) to $\sigma$ and
progressively approach the HS limiting value $0.1 \sigma$.

In a window of $\phi$ values, the crossover between the two plateaux
in the MSD displays a sub-diffusive behavior for up to two decades in
time, i.e.  $\langle r^2\rangle \propto t^{\alpha}$, with a
state-point dependent exponent $\alpha < 1$.  A similar behavior was
found in the simulations of the SW system \cite{Zac02a,Sci03a}, in a
limited $T$-window, within the liquid reentrant region. In the SW case
the subdiffusive behavior is found for MSD values between $\Delta^2$
(the bond cage) and $10^{-2} \sigma^2$ (the HS cage) and it is due to
a competition between attractive and HS glasses at low and high $T$.
Explicit MCT predictions have confirmed this feature\cite{sperl},
connecting it with the presence of a near-by higher order singularity.
In the present case it appears that the subdiffusive behavior arises
from the competition between the very different localization lengths of the
gel-arrested network and the HS glass. The MSD phenomenology is reminiscent
of that found in presence of a higher order MCT
singularity\cite{Gotzesperl}.
%, despite the fact that MCT cannot describe the gel state and it 
%cannot be applied successfully to the present model. 
It is possible that the subdiffusive behavior (and
other features such as the logarithmic decay of the density
auto-correlation functions discussed later on) arise generically from
the competition between two disordered arrested states.

\subsection{Density Relaxation, Bond  Relaxation and Non-ergodicity}

We now focus on the behavior of the density autocorrelation functions
$F_q(t)$. Ref.~\cite{Zaccagel} called attention on the different time
dependence of the low and high $q$ window. 
At small $q$, dynamics slow down 
significantly and become non-ergodic, while at larger $q$ (already on the 
scale of nearest neighbours)  dynamics remain ergodic to within 
numerical accuracy.
%At small $q$ dynamics slows
%down significantly (and becomes non ergodic) while at larger $q$ as in
%the nearest neighbour distance the dynamics remains (within the
%numerical accuracy) ergodic.  
At low density, the non-ergodicity parameter for the gel is different
from that of either the repulsive or attractive glass.  We now
investigate the effect of density on these findings.

\begin{figure}
\hbox to\hsize{\epsfxsize=0.9\hsize\hfil\epsfbox{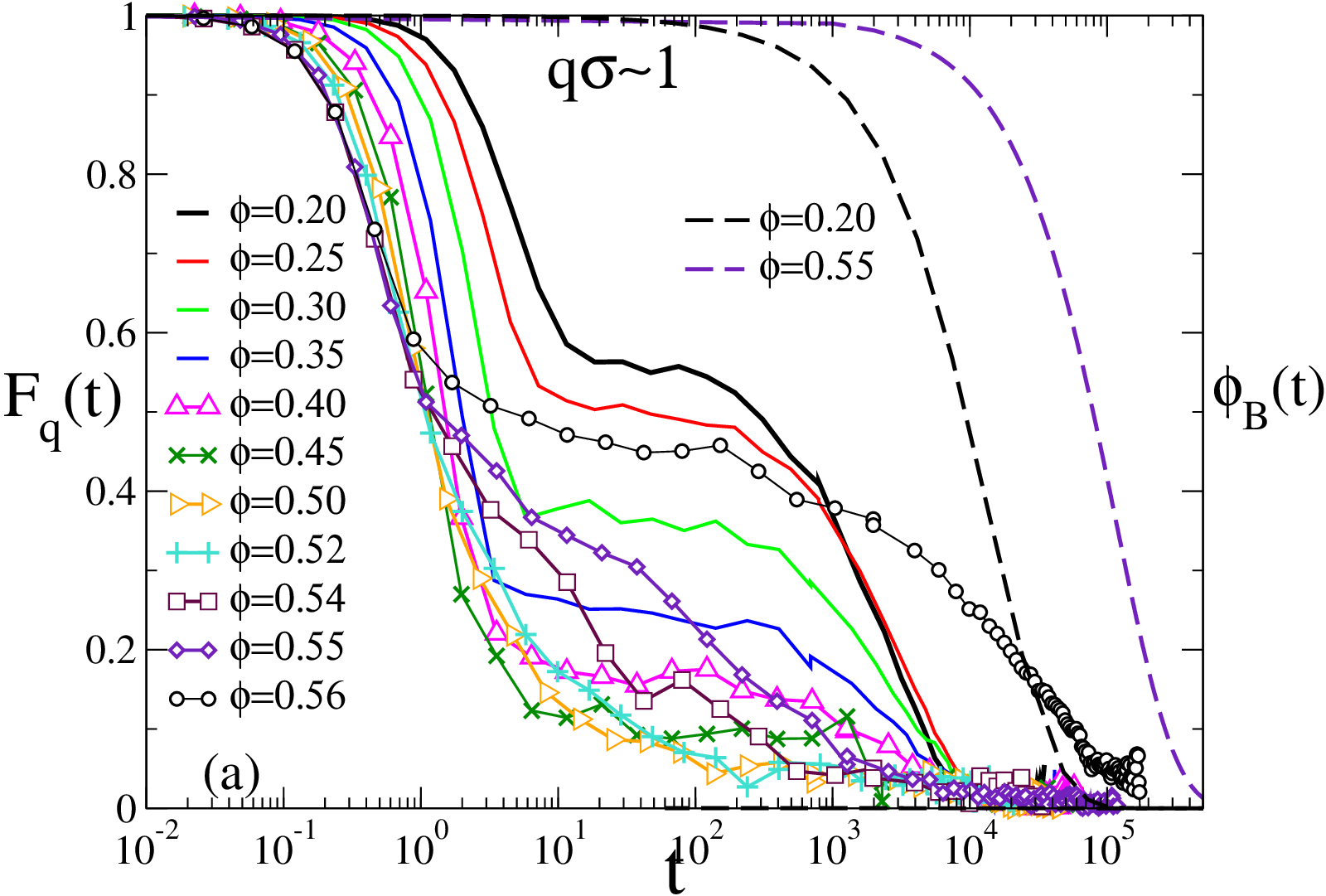}\hfil}
\vspace{0.5cm}
\hbox to\hsize{\epsfxsize=0.9\hsize\hfil\epsfbox{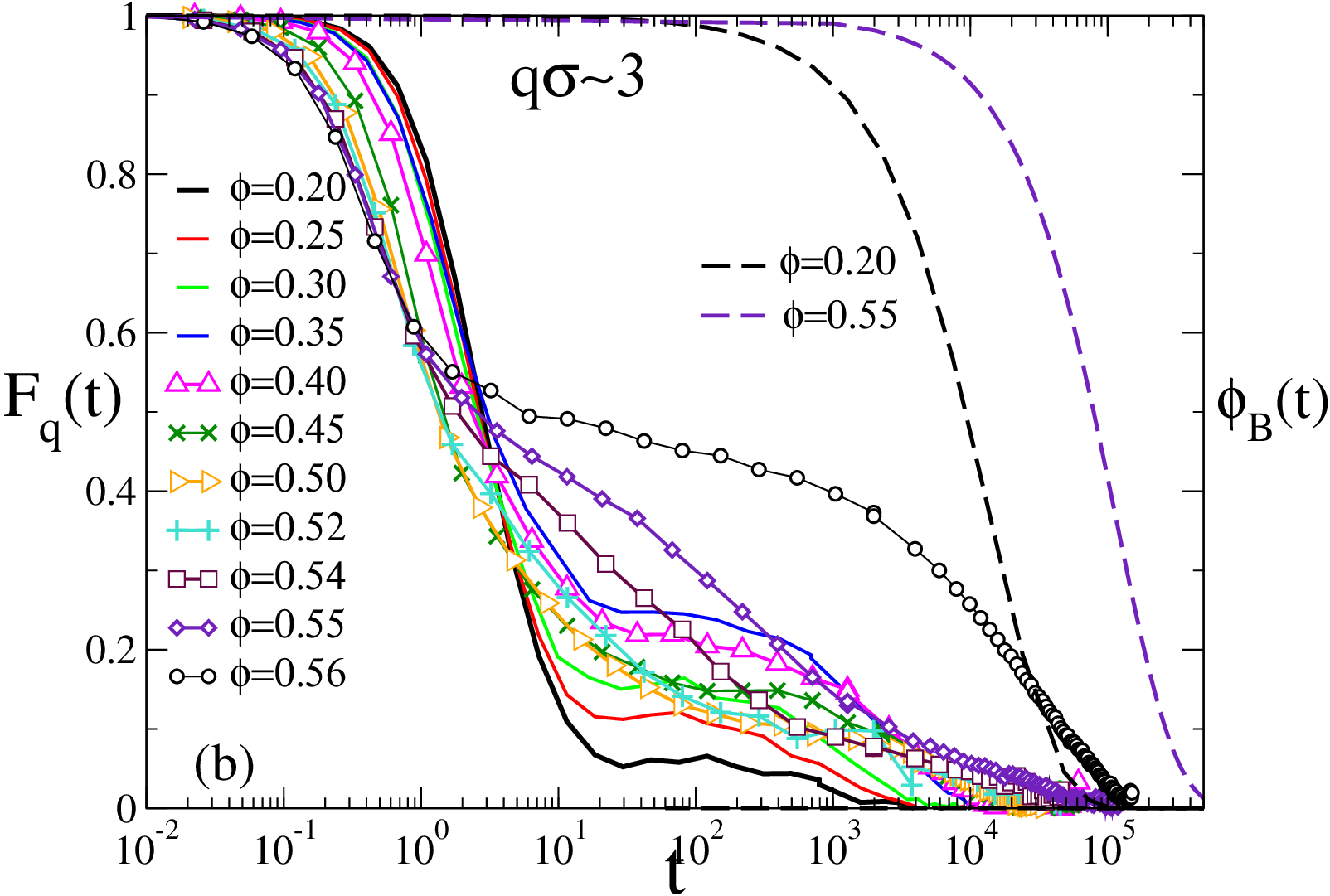}\hfil}
\vspace{0.5cm}
\hbox to\hsize{\epsfxsize=0.9\hsize\hfil\epsfbox{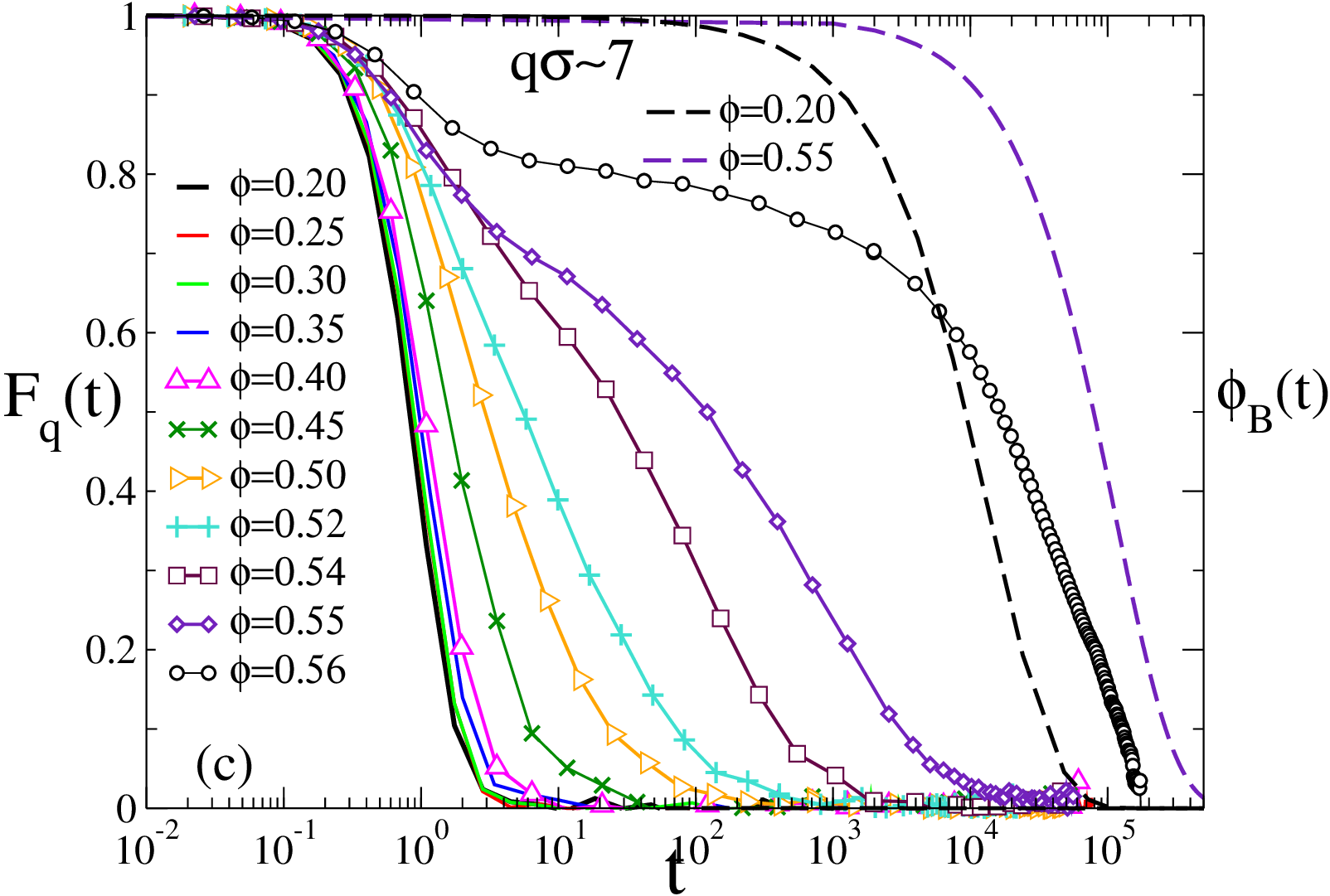}\hfil}
\caption{Density autocorrelation functions for $T=0.1$ and $n_{\rm
max}=3$ at various $\phi$ and $q\sigma\sim 1,3,7$. Also, bond
correlation functions $\phi_B(t)$ (dashes lines) at small and large
$\phi$ are reported in all three panels for comparison. 
}
\label{fig:sqt3}
\end{figure}

To provide a picture of the behavior of the dynamics at low $T$ as a
function of $\phi$ we plot in Fig.~\ref{fig:sqt3} the density
correlators for three different values of $q$, corresponding to
distances respectively much larger, larger, and comparable to the
nearest neighbor distance $\sigma$. In addition, we report the
behaviour of the bond correlation function $\phi_B(t)$ at the same low
$T$, for small and large $\phi$.

The $\phi$-evolution of the shape of the correlation function is
particularly complex. At very small $q$, $(q\sigma\sim 1)$, all
correlation functions show a clear plateau, followed by the
$\alpha$-relaxation process.  Interestingly, the plateau value has a
non-monotonic behavior with $\phi$.  It starts from an high value at
the lowest $\phi$ and decreases down to less than 0.1 before
increasing again on approaching the hard-sphere glass. At the present
time, we have no explanation for these trends.
 
Even more complicated is the $q\sigma\sim 3$ case. The plateau value
first increases with $\phi$, then a reversal of the trend is observed at
$\phi=0.40$ where the plateau height starts to decrease. Such a decrease
persists up to $\phi=0.50$, after which a distinguishable plateau
almost disappears. Correlators are higher at comparable times and
become almost logarithmic for up to three time decades. At
$\phi=0.56$, a clear repulsive glassy behavior is recovered.

Much simpler is the interpretation of the last case, $q\sigma\sim 7$ (and
larger $q$).  Here, the standard scenario for the repulsive glass is
observed, despite the presence of a connected long-living network of
bonds.  The absence of any detectable (within our numerical precision)
plateau at small length scales and low $\phi$ confirms the ability of
the particles to explore distances smaller than $\sigma$ without any
constraint. This is an effect of the loose character of the network,
of the small overall $\phi$ (as compared to the typical HS glass
values) and of the small local degree of connectivity.

Focusing on the behaviour of $\phi_B(t)$, we find that the curves
follow closely, at all studied $\phi$, a simple exponential law,
i.e. a stretched exponential fit gives an exponent $\beta$ always
close to $1$ (tending to $1$ with decreasing $T$). From the figures,
it is evident that bond relaxation is always much slower as compared
to density relaxation even for very small $q$. This suggests that, up
to at least a time of order $10$, the density relaxation is coupled to
the movements of a permanent network which, without breaking most of
its bonds, is capable of spanning a large part of the simulation
box. At longer times, the breaking of the bonds enters into play,
producing a secondary very slow relaxation in $F_q(t)$, accompanied by
very small plateau, as it is evident in panels (a) and (b) for small
$q$. 

\begin{figure}
\hbox to\hsize{\epsfxsize=0.9\hsize\hfil\epsfbox{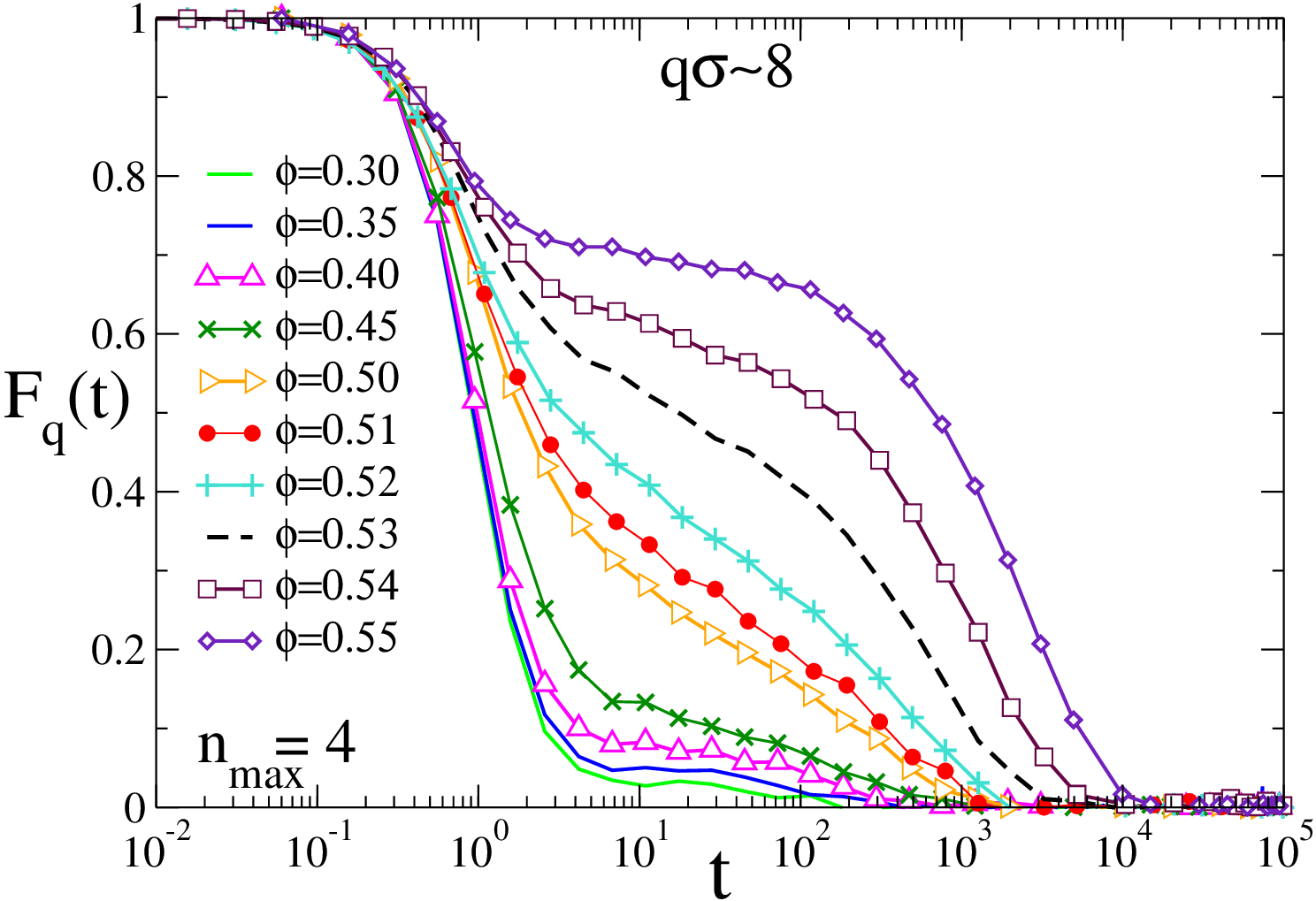}\hfil}
\caption{Density autocorrelation functions for $T=0.125$ 
and $n_{\rm max}=4$ at various $\phi$
and $q\sigma\sim 8$.}
\label{fig:sqtnmax4}
\end{figure}

On increasing $n_{\rm max}$, the same features for $F_q(t)$ are
observed at a progressively larger $q$. An example is reported in
Fig.\ref{fig:sqtnmax4} for the case $q\sigma\sim 8$. Here, around
$\phi=0.52$ a logarithmic decay for about two time decades is
observed. At this wave-vector, corresponding to lengths smaller than
the nearest neighbour one, the non-monotonic behavior of the plateau
is not present anymore.

\begin{figure}
\hbox to\hsize{\epsfxsize=0.9\hsize\hfil\epsfbox{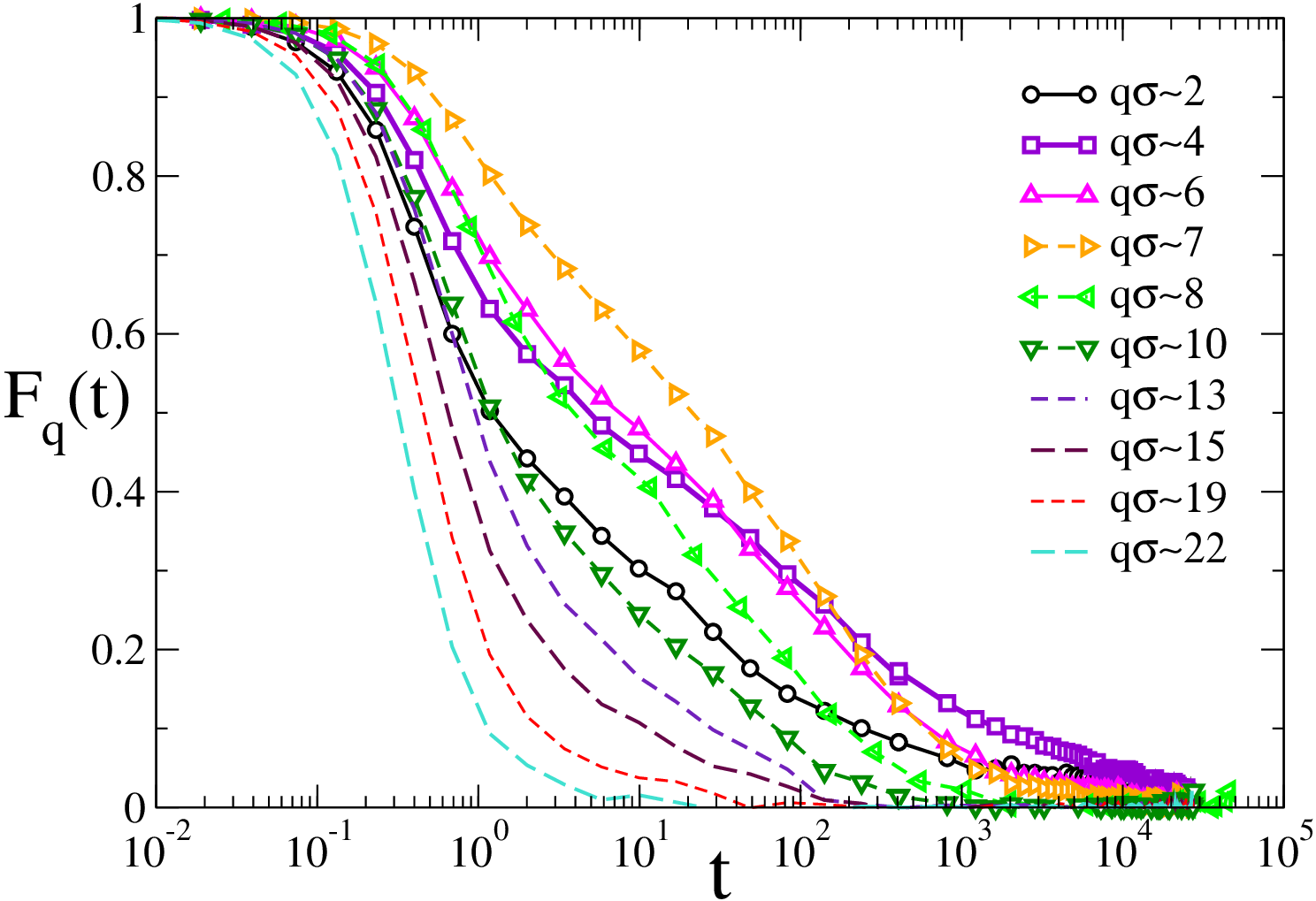}\hfil}
\caption{Density autocorrelation functions for $T=0.1$ , $\phi=0.54$
and $n_{\rm max}=3$ at various $q\sigma$. A $log(t)$ behavior is
observed for $3\leq q\sigma \leq 6$ for up to three time decades.}
\label{fig:sqt-variq}
\end{figure}

To better grasp the logarithmic behavior observed in
Figures.~\ref{fig:sqt3} and \ref{fig:sqtnmax4}, in
Fig.~\ref{fig:sqt-variq} we show the $q$-dependence of $F_q(t)$ at
$\phi=0.54$, i.e. the $\phi$ showing the most enhanced $log(t)$
dependence, for $n_{\rm max}=3$.  We note that the best (long-lasting)
$log(t)$ dependence is seen in a finite window of $q$-values, roughly
between $3 \leq q\sigma \leq 6 $, and it covers about three
orders of magnitude in time. Again, on increasing $n_{\rm max}$ this
$q$-window shifts to larger $q$ (approximately between $8 \leq q\sigma
\leq 10 $).

A possible quantification of the characteristic time of the dynamics
and of the non-ergodicity parameter is provided by stretched
exponential fits ($F_q(t)=f_q \exp{-(t/\tau_{q})^{\beta_q}}$.) of the
long time decay.  This fit allows us to extract information on the
behavior of the non-ergodicity parameter $f_q$, as well as on the
stretching exponent $\beta_q$ and an estimation of the relaxation time
$\tau_q$.  Still, in a certain region of $\phi$ and $q$ values, the
decay of the correlators is clearly different from a stretched
exponential, being $F_q(t)$ essentialy linear in $log(t)$.  Under
these conditions we cannot fit the density correlators and estimate
the non-ergodicity parameter.

\begin{figure}
\hbox to\hsize{\epsfxsize=0.9\hsize\hfil\epsfbox{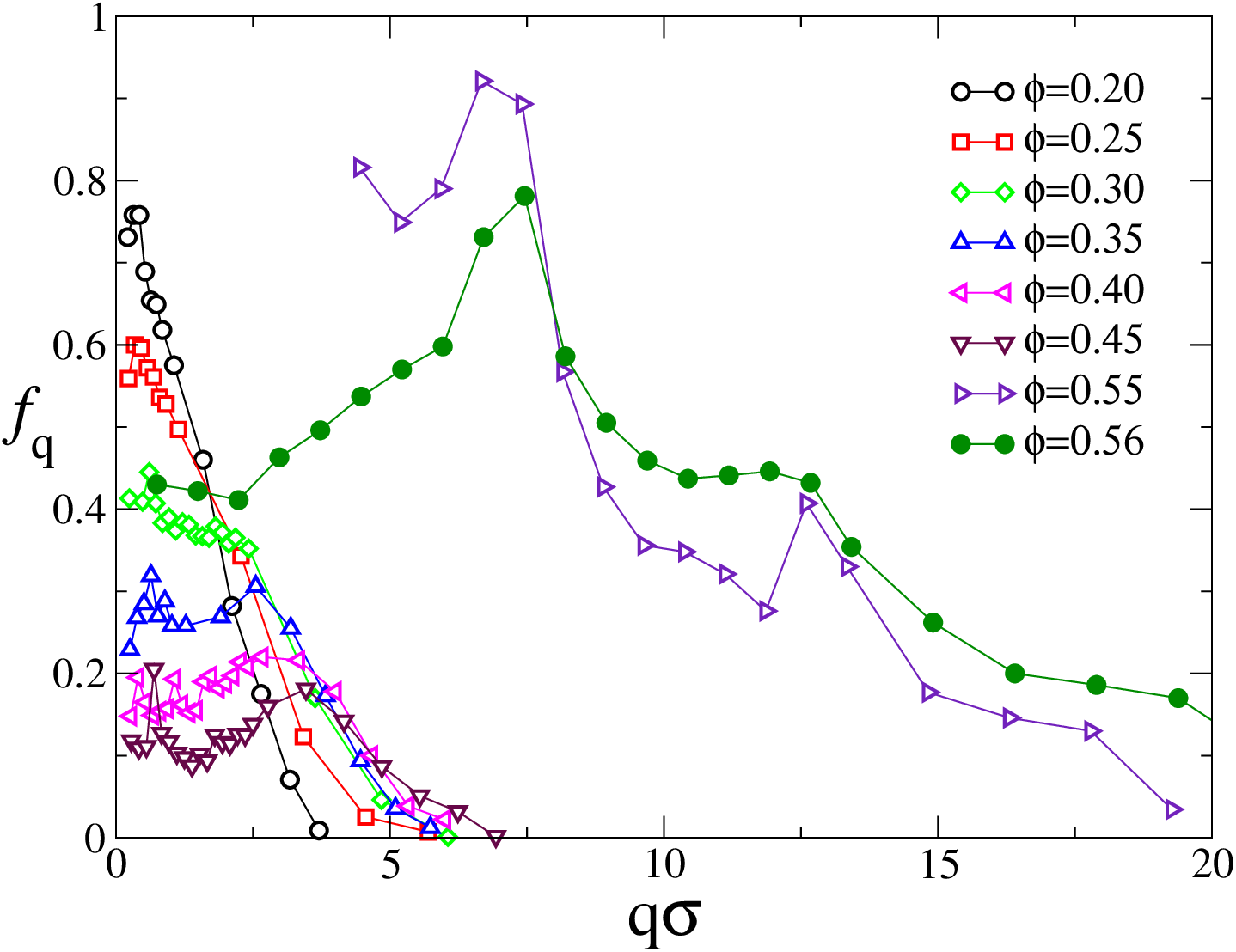}\hfil}
\vspace{0.5cm}
\hbox to\hsize{\epsfxsize=0.9\hsize\hfil\epsfbox{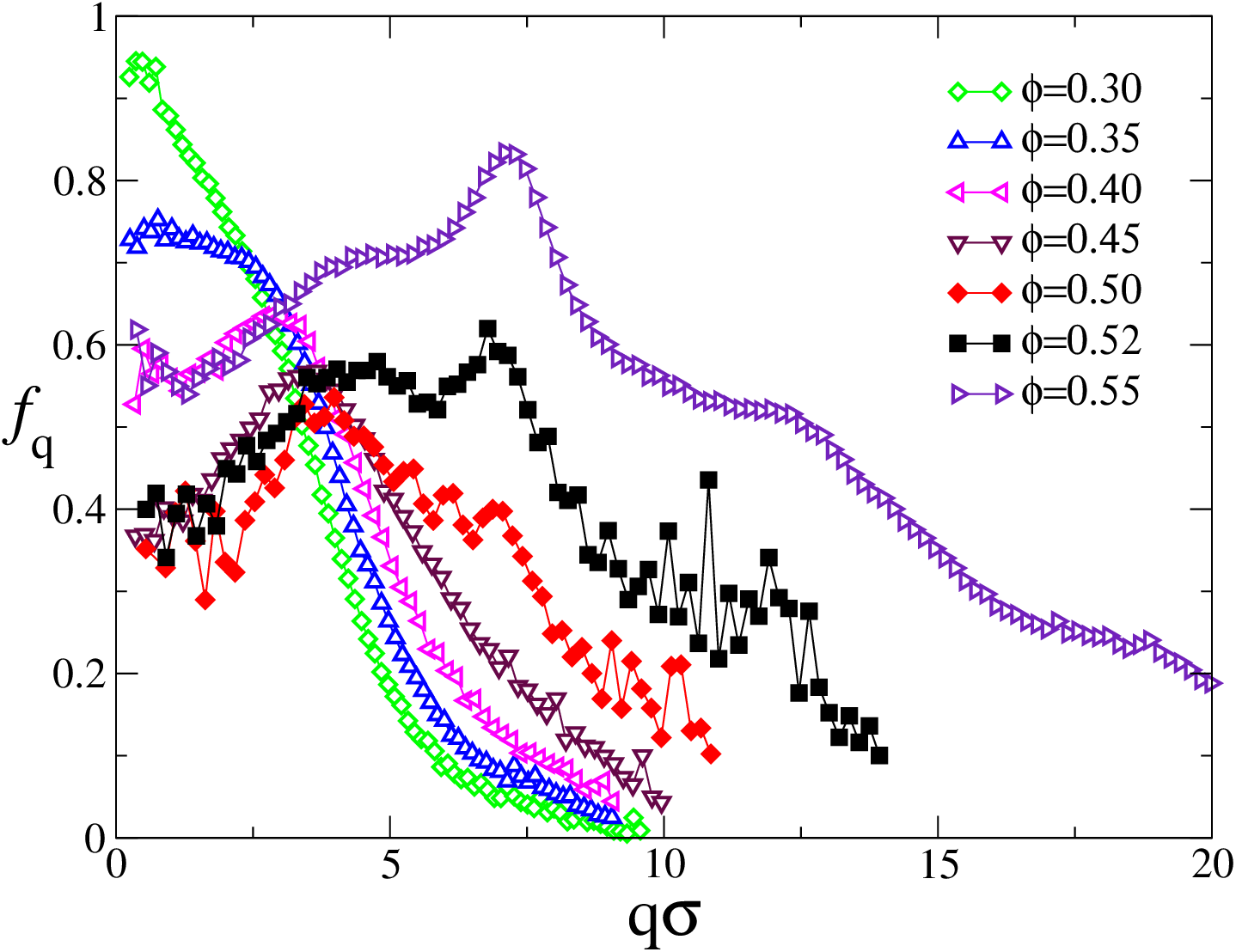}\hfil}
\caption{Non-ergodicity factor, obtained from streched exponential
fits, at various $\phi$ for $T=0.1$ and $n_{\rm max}=3$ (a) and
$T=0.125$ and $n_{\rm max}=4$ (b). }
\label{fig:fqt}
\end{figure}

Fig.~\ref{fig:fqt} shows $f_q$ along the low $T$ isotherm for $n_{\rm
max}=3$ and $4$ for several $\phi$.  In both cases, a different
non-ergodic behavior at low and high $\phi$ is evident.  At low
$\phi$, $f_q$ is largest at $q\rightarrow 0$, then decays rapidly to
zero within a range of about $5$ in units of $q\sigma$.  With
increasing $\phi$, the overall height of $f_q$ decreases, but a small
peak starts to form at larger $q$, which is still of the order of a
few $q\sigma$. This behavior roughly follows that of $S(q)$, for which
the low $q$ increase turns to a small peak at finite $q$ (see for
example in Fig.\ref{fig:sq}b).  At large $\phi$ the shape of $f_q$
closely resembles that of a hard sphere system, with the first peak
around $q\sigma \approx 2\pi$, i.e. the nearest neighbour length
scale.

At intermediate $\phi$, we observe a slightly different behavior
between the two studied values of $n_{\rm max}$.  As discussed before,
for $n_{\rm max}=3$ and $0.45 < \phi < 0.55$, we cannot estimate $f_q$
from the fits.  The reason for this is that $F_q(t)$ displays unusual
features, like a logarithmic decay at certain $q$ followed by a
secondary relaxation reminiscent of the gel type.  Fig.~\ref{fig:fqt}
shows that a sharp transition in the $q$ dependence of $f_q$ takes
place between $0.45 < \phi < 0.55$, which we associate to the
gel-to-glass transition.  For $n_{\rm max}=4$ there seems to be a
smoother transition although a similar non-monotonic behavior of $f_q$
with $\phi$ at low $q$ is observed.

\begin{figure}
\hbox to\hsize{\epsfxsize=0.9\hsize\hfil\epsfbox{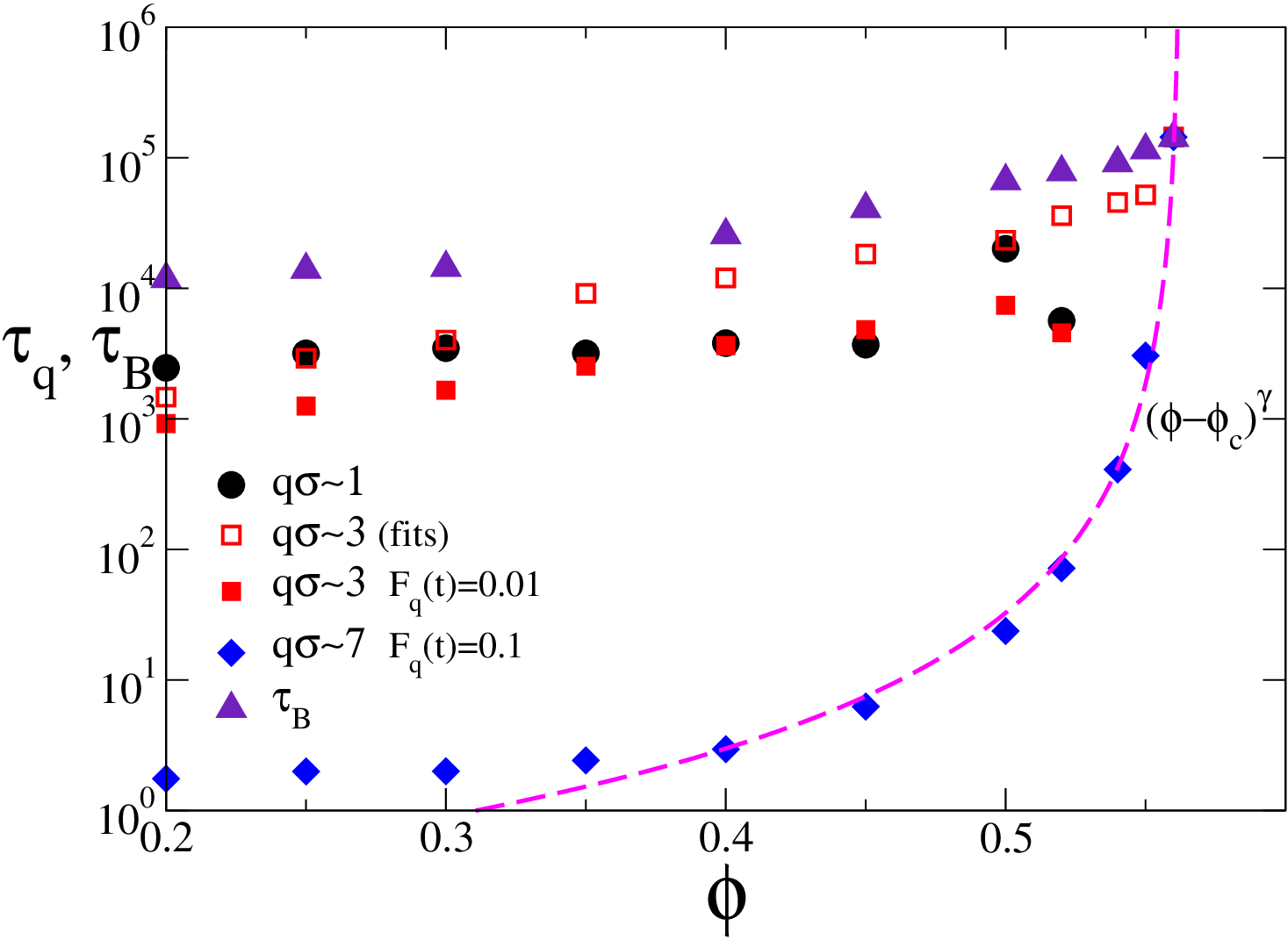}\hfil}
\caption{$\phi$-dependence of the density relaxation time $\tau_q$ at
$q\sigma \sim 1, 3, 7$ versus bond relaxation time $\tau_B$
(triangles), at fixed $T=0.1$ for $n_{\rm max=3}$.  $\tau_q$ are
either extracted from stretched exponential fits or via the relation
$F_q(\tau_q)=0.01$ ($q\sigma\sim 3$) and $F_q(\tau_q)=0.1$
($q\sigma\sim 7$). In the latter case, the dashed line is a fit of the
data to a power law $(\phi-\phi_c)^{\gamma}$.  }
\label{fig:tau_q}
\end{figure}

We further note that the relaxation time $\tau_{q}$, extracted from
the fits or otherwise, monotonically increases with $\phi$ despite the
non-monotonicity in the plateau.  The behavior of $\tau_q$ with $\phi$
at low $T$ is shown in Fig.\ref{fig:tau_q} at the $q$-values discussed
above for $n_{\rm max}=3$, together with the corresponding behavior
for the bond relaxation time $\tau_B$.  Apart from the fits, $\tau_q$
can be also conventionally defined as the time at which the normalized
correlators are equal to an arbitrary (low) value, which is chosen for
convenience as $0.1$.  Doing so, for $q\sigma \sim 7$, we find that
$\tau_q$ follows a typical, for glass-forming systems, power-law
dependence on $\phi$, also shown in the figure, with critical
$\phi_c~0.562$ and exponent $\gamma\approx 2.5$. This result for
$\phi_c$ is quite consistent with that extracted from the diffusivity
fits discussed earlier.  However, on lowering $q$, the situation
becomes more complicated. At $q\sigma \sim 3$, the conventional
arbitrary value $0.1$ is too high, since the correlators displays
smaller plateau values. Thus, in Fig. \ref{fig:tau_q} we report
$\tau_q$ obtained both from the fits and by choosing a conventional
value $0.01$. The results are parallel to each other, and do not show
a power law behavior. Rather there seems to be a crossover regime at
intermediate $\phi$. However, the relaxation time is monotonically
increasing at this length scale of observation, despite the
non-monotonic behavior of the plateau. At even lower $q$,
e.g. $q\sigma \sim 1$, we cannot define a satisfying finite value for
$F_q(t)$, below the plateaux at all studied $\phi$ and numerically
detectable. Also the fits cannot be relied on at high densities. As
shown in the inset, it seems that $\tau_q$ does not vary strongly with
$\phi$, and indeed all correlators (see Fig.~\ref{fig:sqt3}a) seem to
meet at $F_q(t)=0$ at around the same value of $t$, up to $\phi=0.52$.

A sensible estimate of $\tau_B$ is simply obtained by stretched
exponential fits of $\phi_B(t)$.  $\tau_B$ is always longer than the
density relaxation time at all $q$, as remarked above. Moreover, the
increase of $\tau_B$ upon $\phi$, is rather small, indicating that
bonds are slightly stabilized by crowding.  $\tau_B$ is completely
decoupled from $\tau_q$ at large $q$, while it becomes coupled to
$\tau_q$ at small $q$, the data being almost parallel, and, it seems
that in the limit $q\rightarrow 0$, the two relaxation times would
coincide.  Once again, we associate the large $\tau_B$ to the very
slow secondary relaxation observed for density correlators at small
$q$ (see for example $q\sigma \sim 3$ and $0.50 \leq \phi \leq 0.55$
in Fig. \ref{fig:sqt3}b).

A final observation concerns the behavior of $\beta_q$ extracted
from the fits, outside the logarithmic regimes. We find at high $q$
($q\sigma\geq 7$) and low $\phi$, that values of $\beta_q$ are larger
than $1$, actually close to $1.5$, associated more to a compressed
than a stretched exponential \cite{Cip00a,Cip05a,DelG05b}.  However, this
value decreases below $1$ for high $\phi$ and large $q$, while it
tends to be $1$ for low $\phi$ and low $q$ ($q\sigma \lesssim 7$). 
We note that care has to be taken when exponents greater than $1$ are 
found in Newtonian dynamics, since they could arise from undamped 
motion of clusters (possibly of rotational origin), or elastic motion 
within the percolating structure.  A comparison with a
Brownian dynamics simulation may help to clarify this issue.

\begin{figure}
\hbox to\hsize{\epsfxsize=0.9\hsize\hfil\epsfbox{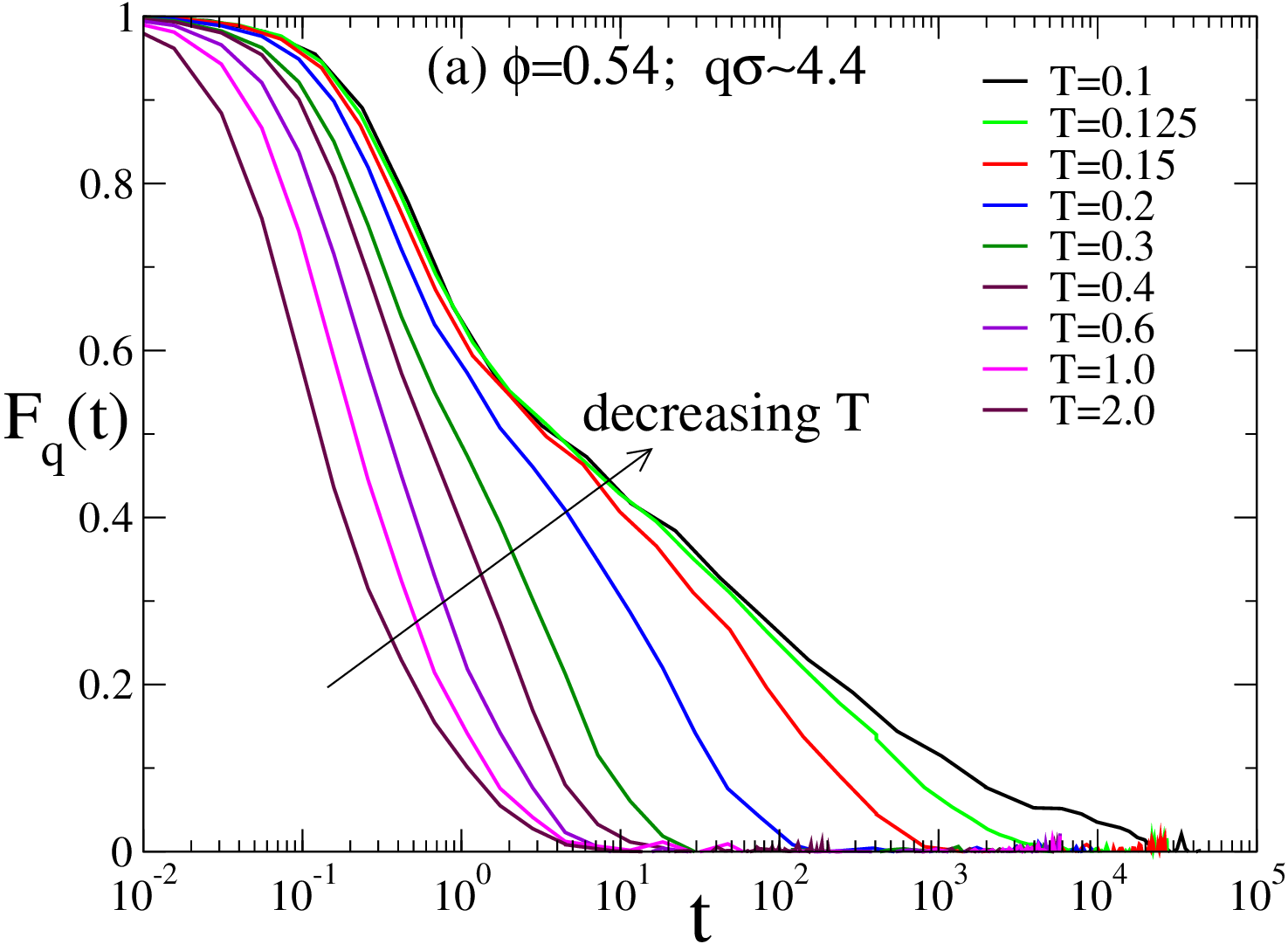}\hfil}
\vspace{0.6cm}
\hbox to\hsize{\epsfxsize=0.9\hsize\hfil\epsfbox{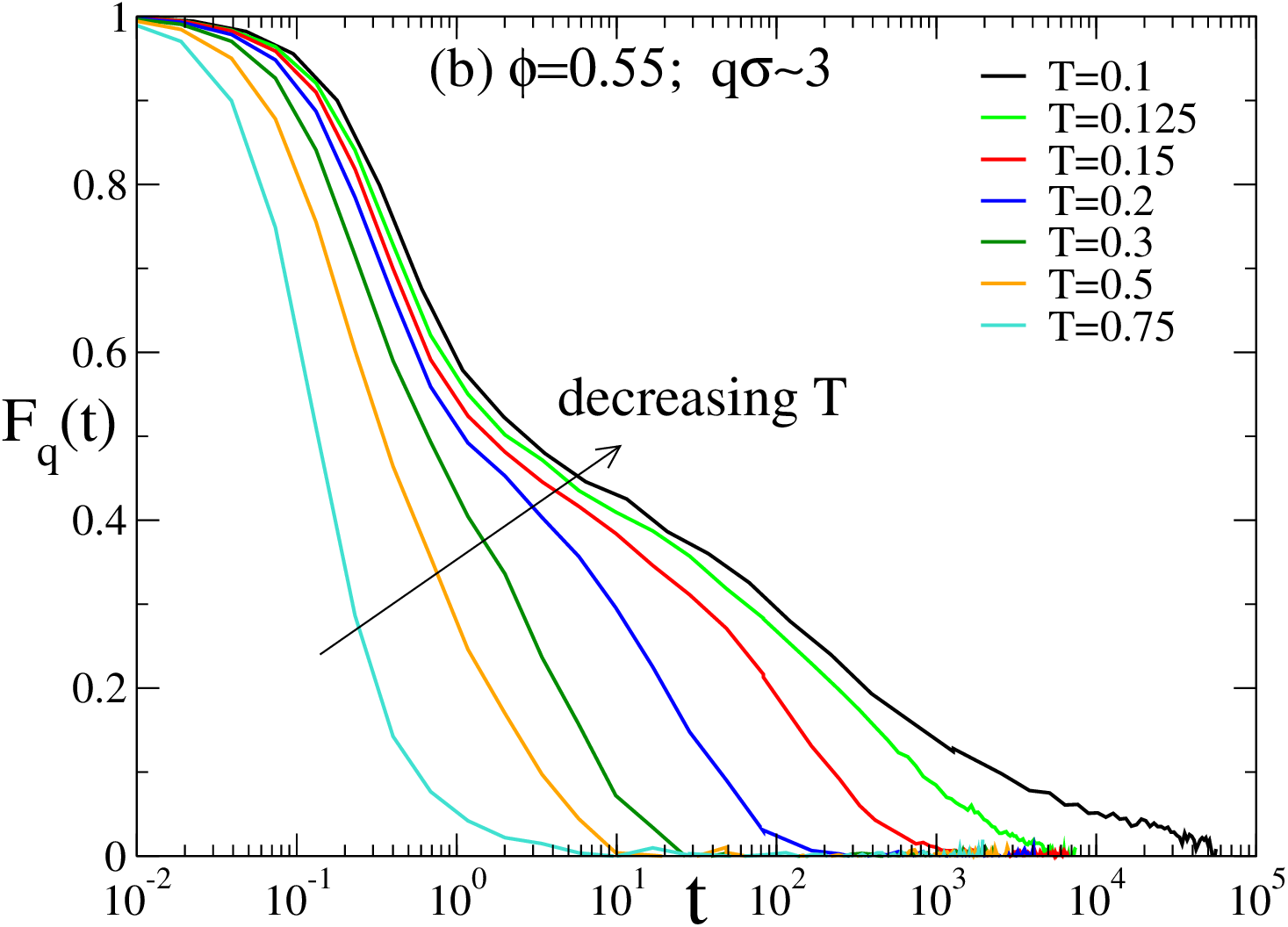}\hfil}
\vspace{0.3cm}
\caption{Density autocorrelation functions for $n_{\rm max}=3$ at
various studied $T$ for (a) $\phi=0.54$ and $q\sigma\sim 4.4$
(maximally enhanhced $log(t)$ behavior) and (b) $\phi=0.55$ and
$q\sigma\sim 3$ (interference between $log(t)$ behavior and standard
glass-like $\alpha$-relaxation). }
\label{fig:sqt-T}
\end{figure}

Additionally, we focus on the $T$ dependence of the density correlators.
In Ref.\cite{Zaccagel}, we discussed the $T$ dependence along the
lowest isochore $\phi=0.20$. There, we found that, only at low enough
$q$, there was the emergence of a gel plateau at low $T$ and that, in
this region of wavevectors, the density relaxation time $\tau_q$ as
well as the bond relaxation time $\tau_B$ both followed an Arrhenius
law. On increasing density, there is the gel to glass crossover, which
is clearly visible by looking at the $T$ behavior of $F_q(t)$, as
shown in Fig. \ref{fig:sqt-T}. Up to $\phi=0.45$ a clear gel plateau
is approached. Beyond this value, the anomalously slow logarithmic
decay is observed, at its best for $\phi=0.54$, reported in figure
\ref{fig:sqt-T}(a). Above this $\phi$, e.g. $\phi=0.55$ in
\ref{fig:sqt-T}(b), a crossover from logarithmic to standard glass
regime (i.e., typical two-step relaxation) is observed.  A clear
difference between these two graphs is the fact that at $\phi=0.54$
pure logarithmic decay is observed up to the lowest $T$, while at
$\phi=0.55$ a kink, evidence of the nearby glass transition, is
present.  We also note that, interestingly, the shape of the
correlators in Fig.  \ref{fig:sqt-T}(b) is very reminiscent of Fig.11
in Ref.\cite{Daw00a} and Fig.6 in Ref.\cite{Zac02a}, respectively MCT
predictions and simulation results for the SW model, for fixed
temperature in the reentrant region and varying density. Here, the
analogous plot is reported, at a density within the gel to glass
crossover, and varying $T$.  Again, this suggests a close similarity
of the features at the gel to glass crossover with respect to the SW
crossover from repulsive to attractive glass.

Finally, both the relaxation time $\tau_q$ and the bond
relaxation time $\tau_B$ are found to obey Arrhenius dynamics in $T$
at all $\phi$.

\section{Discussion} 

The interpretation of these results is not completely straightforward.
However, they strongly suggest that the system gels at low $\phi$ and
forms a glass at high ones.  It is important to understand what
happens in between these two regimes, where a new type of relaxation
takes place, as a result of the competition of the two
effects. Similarly to the simple SW case, where anomalous dynamics
arises in the reentrant liquid region from the competition between
attractive and repulsive glass, here again anomalous dynamics is
generated from the presence of two distinct arrested states. However,
in the attractive-vs-repulsive glass scenario, temperature is the
control parameter generating a liquid pocket in between. Here $\phi$
is the control parameter. We have to bear in mind that at $T=0.1$ the
energy per particlce of the system for $n_{\rm max}=3$ varies from
about $-1.48$ at $\phi=0.20$ to about $-1.495$ at $\phi=0.56$, indicating
that most of the particles ($\approx 99\%$) are fully bonded
(i.e. they have reached the $n_{\rm max}$ limit already). With
increasing $\phi$, more neighbors surround each particle but only
$n_{\rm max}$ of them interact via an attractive well, the others
probing only the hard core interaction. Thus we cannot expect a
non-monotonic $\phi$ dependence of the characteristic time (i.e. no
re-entrance).

Where does the logarithm/subdiffusivity come from?  A more intuitive
understanding of the anomalous dynamics results from interpreting the
MSD behavior.  If one thinks simply of a filling up of space, the MSD
plateau should monotonically decrease and no subdiffusive behavior
should be observed.  However, at $\phi\approx 0.54$ a clear $t^{\alpha}$
law is observed.  After the ballistic regime, particles start to feel
the presence of the nearest-neighbours and the MSD slows down. At long
times, particles are able to break and reform the bonds, the network
fully restructures itself and proper diffusion is observed. In the
intermediate time window one observes the competition between
excluded-volume confinement and exploration of space associated to the
motion of the unbroken network.  We believe that this is at the origin
of the anomalously slow diffusion and logarithmic decay.

To support this hypothesis we note that the logarithmic decay shows up
only in a window of small $q$'s significantly less than the
nearest-neighbour inverse length (i.e. over distances where
connectivity is probed).  On increasing $n_{\rm max}$, this length becomes
smaller due to the higher degree of constraint for the network, in the
same way as the localization length, estimaed from the MSD, decreases.
However, we recall that in the SW system such anomalous $log(t)$ decay
was observed for very large $q$, associated to the typical distance of
the short-range attraction. This provides further evidence that in the
$n_{\rm max}$ case the connectivity of the network is associated with 
generating unusual $log(t)$ features and confirming that the width of
the attractive well does not play any significant role.

A final comment regards the existence of the so-called higher order
MCT singularity. In the SW case, the width of the attraction $\Delta$
is the crucial control parameter driving the system close to the
singularity. In the $n_{\rm max}$ case, $\Delta$ does not play a
relevant role. It is intriguing to ask ourselves which parameter plays
the role of $\Delta$, if the gel-to-glass cross over belongs to the
same class of models possessing a higher order MCT singularity.  One
possible answer is $n_{\rm max}$ itself. Indeed, we noted how the
characteristic length scale to observe logarithmic behavior in
$F_q(t)$ shifts with increasing $n_{\max}$. Our present knowledge is
that this is found at $q\sigma \approx 4$ for $n_{\max}=3$ and
$q\sigma \approx 8$ for $n_{\max}=4 $. We also know that for the SW,
although being associated to the competition of repulsive and
attractive glasses, it moves up to $q\sigma \sim 20$\cite{Sci03a}. The
possible existence of a smooth crossover between the two phenomena
could be theoretically investigated with simulations. However, the
existence of the higher order singularity in the present model is
destined to be uncertain, unless some theory is devised for the gel
transition and its predictions tested against the simulations.

\section{Conclusions}

Understanding the slowing down of the dynamics in colloidal systems
and the loci of dynamic arrest in the full $(\phi-T)$ plane,
encompassing gel and glass transitions, is one of the open issues in
soft condensed matter.  Two classes of potentials have been explored
in some details in the recent years: (i) short-range attractive
spherical potentials and (ii) short-range attractive spherical
potentials complemented with a repulsive shoulder.  In the first case,
it has been shown that low $\phi$ arrested states arise only as a
result of a interrupted phase separation.  The second case appears to
be much more complicated and partially unresolved. For certain values
of the parameters of the repulsive potential the system separates into
clusters (which can be interpreted as a mesoscopic interrupted phase
separation) and dynamic arrest at low $\phi$ can follow from a cluster
glass transition or from cluster percolation.

In the attempt to provide an accurate picture of dynamics and a model
for dynamic arrest at low $\phi$ in the absence of phase separation
(both at the macroscopic and mesoscopic level) we present here
a study of a minimal model of gel-forming systems. The model
builds up on the intuition that phase separation is suppressed when
the number of interacting neighbors becomes less than six, since the
energetic driving force for phase separation becomes less
effective\cite{sastry}.  To retain the spherical aspect of the
potential, a standard SW interaction potential is complemented by a
constraint on the maximum number $n_{\rm max}$ of bonded neighbors (a
model similar to the one first introduced by Speedy and
Debenedetti\cite{Spe94,Spe96}).

In this manuscript we have presented a detailed study in the
$(\phi,T)$ plane of the dynamics for  $ n_{\rm max}=3$ and $n_{\rm
max}=4$. For these two values of $n_{\rm max}$ the region of phase
diagram where unstable states (with respect to phase separation) are
present shrinks to $T \lesssim 0.1$ and $\phi \lesssim 0.3$, making it
possible to approach on cooling low $\phi$ arrested states,
technically in metastable (with respect to crystallization)
equilibrium.  
%General comment, how do we know that the crystal has the lowest free 
%energy?  We might technically not be in metastable equilibrium.    What 
%does the last sentence mean: It becomes possible to predict... 
%kinetically stabilized as compared to the lowest free energy 
%crystalline states?
The simplicity of the model makes it possible to study
it numerically even at very low $T$ and estimate, with accuracy, the
low-$T$ fate of the supercooled liquid. It becomes possible to predict
the regions in the phase diagram where disordered arrested states are
kinetically stabilized as compared to the lowest free energy
crystalline states.

Dynamics in the $n_{\rm max}$ model is also important because it
provides a zero-th order reference system for the dynamics of
particles interacting via directional potentials.  These systems
include globular protein solutions (hydrophilic/hydrophobic patches on
the surfaces of proteins), the new generation of patchy colloids, and,
at a smaller scale, network forming liquids.  In this respect, the
$n_{\rm max}$ model allows us to study the generic features (since it
neglects the geometric correlations induced by directional forces) of
particle association. It has the potential to provide us with an
important reference frame to understand dynamical arrest in
network-forming liquids and the dependence of the general dynamic and
thermodynamic features on the number of patchy interactions.

One of the important results is contained in Fig.~\ref{fig:perc_spin},
which shows the locus of iso-diffusivity in the $(\phi-T)$ plane.
These lines, which provide an accurate estimate of the glass
transition line ($D \rightarrow 0$) are found to be essentially 
vertical at high $\phi$, in correspondence to the HS glass transition, 
and essentially  horizontal at low $T$. Only a very weak, almost 
negliglible, reentrance in $\phi$ is observed. Extrapolating the 
$D$-dependence by power laws in $\phi$  and by Arrhenius laws in $T$, we 
estimate  the glass-lines.    The Arrhenius law is found to be valid 
for many decades, suggesting that, at intermediate $\phi$, $D$ vanishes 
only at $T=0$ . Data strongly support the possibility of two 
distinct arrest transitions:  a glass of the HS type driven only by 
packing at high $\phi$ and a gel at low $T$ \cite{notaTgel}.  
%\footnote{Of course, for 
%convenience,
%in analogy to the definition of the glass transition, one can define a
%timescale above which the system is considered to be gelled, and use
%that as the definition of $T_{gel}$. In any case, independently of
%this definition, the gel line is rather flat.  }

We have chosen the word {\it gel} to label arrest at small and
intermediate $\phi$ since the analysis of the simulation data confirms
that the establishment of a network of bonded particles and the
network connectivity plays a significant role in the arrest process.
While particles are locally caged by SW bonds with $n_{\rm max}$
neighbors (and in this respect one would be tempted to name it an
attractive glass), particle localization is not only controlled by
bonding. Bonded particles are free to explore space (retaining their
connectivity) until they are limited by the network constraints.
Indeed, the plateau of the MSD is, especially at the lowest $\phi$,
larger than the particle size.  We remark that the bond localization,
typical of the attractive glass case, is not observed throughout the
phase diagram, neither in the MSD nor in the width of $f_q$. We
believe this is due to the fact that, although bonding is present,
particles are confined by the potential well only relative to each
other.  The network connectivity length is the quantity that enters
into the determination of the localization length in the arrested
state.  On increasing $\phi$, the localization length progressively
approaches the one characteristic of the hard-sphere glass, signaling
that a cross-over to the excluded volume case takes place.

The intersection between the repulsive glass and gel loci appears to
be associated to anomalous dynamics.  No intermediate liquid state,
i.e. no reentrant regime in $\phi$, is found.  Interestingly enough,
these anomalies are strongly reminiscent of the anomalies observed
close to the intersection of the attractive and repulsive glasses in
the case of short-range interacting particles.  Correlation functions
show a clear $\log(t)$ dependence in a window of $q$ vectors and the
MSD shows a clear subdiffusive behavior $\sim t^{\alpha}$.  Here, the
gel localization length is larger that of the HS glass, a different
scenario from the attractive-repulsive case.  These results support
the hypothesis that a possible MCT-type higher order singularity in
the $n_{\rm max}$ model is present and, at the same time, provide
further support to the intrinsic difference in the localization
mechanisms that are active for the two arrested states.  In contrast
to the SW case, the well-width is not a crucial length scale in the
problem, while an important parameter is $n_{\rm max}$, that could be
the control parameter of a putative higher order singularity of the
MCT type.

%% Using the numeric "exact" $S(q)$, we have solved the MCT equations and
%% evaluated the glass lines.  The theoretical results differ only
%% slightly from the SW case and do not provide any indication of a gel
%% line. According to the theory an attractive glass line (i.e. with the
%% typical features of the attractive glass as MSD $\sim \Delta^2$ and
%% very wide $f_q$) should be present. From a theoretical point of view,
%% such a line arises from the large $q$ oscillations in $S(q)$,
%% i.e. from the $q$-space signature of the short-range bonding, i.e. the
%% same origin as in the SW.

%Indeed, at low $\phi$, oscillations
%of parts of the connected network are possible, preventing the
%observation of the MCT mechanism for arrest.  

%% Of course, the inclusion in the MCT calculation of higher order
%% correlation functions could be important in the study of a many-body
%% interaction and should be considered. However, due to the sphericity
%% of the model, our opinion is that this should not make a significant
%% change, i.e. the prediction for the attractive glass transition should
%% be robust in this respect.

One further consideration refers to the role of $\Delta$ in the
$n_{\rm max}$ model.  We have chosen to use $\epsilon \equiv
\Delta/(\sigma+\Delta)=0.03$ to connect with the well studied
corresponding SW case. We do not expect significant differences for
the two arrested states for $\Delta$ values up to $\epsilon \approx
0.1-0.2$, since the connectivity properties would be essentially
identical. Thus, we still expect the existence of distinct gel and
glass lines. Only the interplay between the two could be affected, as
for example the $log(t)$ behavior should be shifted in its
$q$-dependence. Also, in that case, the behavior with $n_{\rm max}$ is
not a priori clear, since for larger $\Delta$ values no MCT
singularity is present for the SW model and, from a theoretical point
of view, the attractive glass line is not physically distinct from the
repulsive one.

In summary, the present model provides a clear indication that even if
liquid-gas phase separation can be avoided and arrest at low
$\phi$ can be explored in equilibrium conditions, the observed
arrested state is not the low-$\phi$ extension of the attractive
glass.  
%The comparison with the solution of the MCT equations for the
%$n_{\rm max}$ model (a possibility offered by the spherical symmetry
%of the interaction potential) confirms that MCT significantly
%overestimates the role of the bonding, predicting an attractive glass
%even in the present case.  
The present results strongly suggest that
the attractive glass is an arrested state of matter which can be
observed in short-range attractive potentials only at relatively high
$\phi$, being limited by the spinodal curve.  When the inter-particle
potential favors a limited valency, arrest at low $\phi$ becomes
possible but with a mechanism based on the connectivity properties of
a stable particle network, clearly different from what would be the
extension of the (attractive) glass line.

\section{Acknowledgments}
We acknowledge support from MIUR-Cofin, MIUR-Firb and
MRTN-CT-2003-504712.  I.~S.-V. acknowledges NSERC (Canada) for
funding.  We thank T. Voigtmann and W. Kob for useful discussions.

\bibliographystyle{./phaip}
\bibliography{./articoli,./altra}

\begin{thebibliography}{10}
\bibitem{notaTgel}
Of course, for convenience, in analogy to the definition of the glass
  transition, one can define a timescale above which the system is considered
  to be gelled, and use that as the definition of $T_{gel}$. In any case,
  independently of this definition, the gel line is rather flat.
\end{thebibliography}

\end{document}